\documentclass[preprint, 3p, times, sort&compress]{elsarticle}

\usepackage{amsfonts, amsmath, amssymb}
\numberwithin{equation}{section}

\usepackage[hidelinks, pdfauthor=author]{hyperref}

\usepackage{verbatim, xcolor}
\usepackage{tabularx, enumitem, epsfig, float, subcaption, cleveref}

\usepackage{pgfplots, tikz, tikzscale}
\usetikzlibrary{decorations, decorations.pathreplacing}
\pgfplotsset{compat=1.14}

\newcommand{\subf}[2]{{\small \begin{tabular}[t]{@{}c@{}} #1\\#2 \end{tabular}}}

\let\vec\boldsymbol

\begin{document}
\begin{frontmatter}

\journal{}

\title{Shape-shifting panel from 3d-printed undulated ribbon lattice}

\author[address1]{Filippo Agnelli}
\author[address2]{Michele Tricarico}
\author[address1]{Andrei Constantinescu\corref{mycorrespondingauthor}}
\cortext[mycorrespondingauthor]{Corresponding author}
\ead{andrei.constantinescu@polytechnique.edu}

\address[address1]{Laboratoire de Mécanique des Solides, CNRS, École polytechnique, Institut polytechnique de Paris, 91128 Palaiseau, France}
\address[address2]{MMC Laboratory, Department of Engineering Science, University of Oxford, Parks Road, Oxford, OX1 3PJ, United Kingdom}

\date{Pre-print: \today}

\begin{abstract}
Materials that change their shape in response to external stimuli opens up new prospects for efficient and versatile design and shaping of three-dimensional objects. Here, we present a novel class of micro-structures exhibiting an extension-bending coupling (EBC) effect, that can be harnessed as an elementary building block for shape-shifting panels. They are built with a single material as a network of undulated ribbons. The deformations mechanisms of both single and connected undulated ribbons are analysed using the finite element method to explain the main features of the EBC mechanism. For a particular micro-structure of the proposed class, the complete elastic stiffness tensor is computed combining two-scale homogenization with Kirchhoff-Love plate theory. The range of achievable EBC ratio is then assessed with respect to the geometric parameters of the unit cell. Patterned specimens are manufactured using a commercial FFF Ultimaker 3-d printer and are mechanically tested at finite strain up to $20\%$. The displacement measured by point tracking match the predictions from the finite element simulations and indicate that the structure maintain its properties at finite strain. Moreover, a tensile test load with point-like boundary is proposed to highlight exceptional out of plane displacement. We envision these structures to be leveraged in combination with responsive materials for the actuation of soft robots, compliant systems and reconfigurable structures, as alternatives to external mechanical motors, control systems and power devices.
\end{abstract}

\end{frontmatter}

\section{Introduction}
The morphing of shell-based structures into programmable three-dimensional geometries is a ubiquitous mechanism found in nature, which is attracting increasing interest for technological applications \cite{Oliver2016}. In engineering, flat panels are traditionally enticing due to their high strength-to-weight ratio which makes them structurally efficient. Furthermore, they are easily fabricated (\textit{e.g.} by punching, water jetting, laser cutting) and suitable for storage and transportation \cite{Malomo2018, Panetta2019}. Including programmability to such structures expands the range of available manufacturing techniques and increase the fabrication throughput for three-dimensional objects of complex geometries \cite{Na2014, Khoo2015}. In addition, shape-shifting systems unleash new functionalities suitable for exploring harsh or inaccessible environments \cite{Rafsanjani2018, Kotikian2019} and delivering increasingly large and complex payloads \cite{Bauhofer2017, Chen2018}. Several concepts are already reported in minimally invasive surgery \cite{Randall2012, Cianchetti2014}, automotive \cite{Daynes2013}, aeronautics \cite{Ajaj2016} and up to space sector, through antennas \cite{Jacobs2012} and space-based solar power \cite{Arya2016, Chen2019}.\smallskip

Systems with shape changing capacities are obtained through ``mechanisms'' occurring in the micro-architecture of the material. Current digital manufacturing technologies such as 3-d printing \cite{MacDonald2016, Truby2016, Kuang2018} and laser cutting \cite{Shan2015a, Tang2017, Mizzi2020} offer a wide the range of achievable micro-architectures that couple locally prescribed in-plane kinematics to changes in curvature. Mechanisms yielding coupled deformation modes are achieved imparting deformable sheets with cellular micro-architectures, cut patterns or folding patterns. Origami structures for instance may be turned into nearly arbitrary shapes \cite{Wei2013, Overvelde2016}. Yet, due to the independent folding motions of individual folds, they are challenging to fold \cite{Demaine2017} or actuate, relying on complex non-mechanical stimuli \cite{Felton2014, Mao2015, Plucinsky2018}. Kirigami tessellations allow compact flat shapes to conform approximately to any prescribed target shape in two or three dimensions \cite{Konakovic2016, Choi2019}. Shape-shifting concepts have been demonstrated combining folding and cut patterns in \cite{Neville2016}, and exploiting geometric frustration induced in non-periodic cut patterns in \cite{Celli2018}. They are suitable to shape three-dimensional objects with desired geometrical sizes and aspects ratios, but due to the thin connections at the hinges, they are mechanically weak. Ribbon- and membrane-like flat micro-structures \cite{Chopin2013} can buckle out of plane and produce three-dimensional geometries when subject to mechanical actuation \cite{Xu2015, Celli2020}. Lastly, bilayers sheets \cite{Gude2006, vanRees2017} can morph into three-dimensional surfaces with non-zero Gaussian curvature, but their fabrication is complex.\smallskip

Most systems are paired with mechanical actuations through manual forming, boundary loading, or through the release of a pre-stretched layers. By releasing pre-stretched shape memory layers, it is possible to control the deformation in time, which is an essential feature to prevent collisions while undertaking complex morphing \cite{Mao2015, Guseinov2020}. Other studies make use of pneumatic power to mechanically load the shapes \cite{Rafsanjani2018, Siefert2018}. Alternatively, combining shape-morphing structures with multiphysics phenomena further opens up the space for various actuation mechanism. Self-actuation enables autonomous structural adaptation to changing environmental stimuli. For example, self-shaping concepts have been demonstrated in shells through hydrogel swelling \cite{Gladman2016} and nematic-to-isotropic phase changes in liquid-crystal elastomers \cite{Kotikian2018}. Multiple materials in heterogeneous lattice designs enabled unprecedented morphing capacities with complex and doubly curved shapes (\textit{e.g.} a human face) \cite{Boley2019}. Nevertheless, complex shapes remain difficult to achieve experimentally, as they often require advanced multi-material 3-d printers with limited and costly fabrication.\smallskip

In this work, we propose novel a class of micro-structures consisting in a combination of undulated ribbons, parametrised using b-spline surfaces. The undulations feature an asymmetry along the height that is leveraged to obtain an extension-bending coupling (EBC) mechanism. The unit cell is tessellated periodically to generate panels with programmable morphing capabilities when subject to mechanical actuation. While mere undulated ribbons do not exhibit specific coupling mechanics, we demonstrate that their interconnection starts the mechanism. We then discuss the mechanical properties of a particular of a unit cell, computing the complete elastic stiffness tensor via two-scale homogenization with thin plate theory. The range of achievable EBC ratio is then assessed with respect to the geometric parameters of the unit cell. Specimen of the panels are manufactured using a desktop fused filament fabrication 3-d printer and are mechanically tested for validations. Both experiments and numerical simulations are conducted to measure the out of plane local mechanical fields. Our work distinguishes itself for the simplicity of fabrication and actuation, and for its potential applicability in material and structural systems at vastly different scales; it therefore illustrates a potential base to be harnessed in combination with responsive materials for the actuation of soft robots, compliant systems and reconfigurable structures, as alternatives to external mechanical motors, control systems and power devices.

\begin{figure}
\centering
\includegraphics[scale=1]{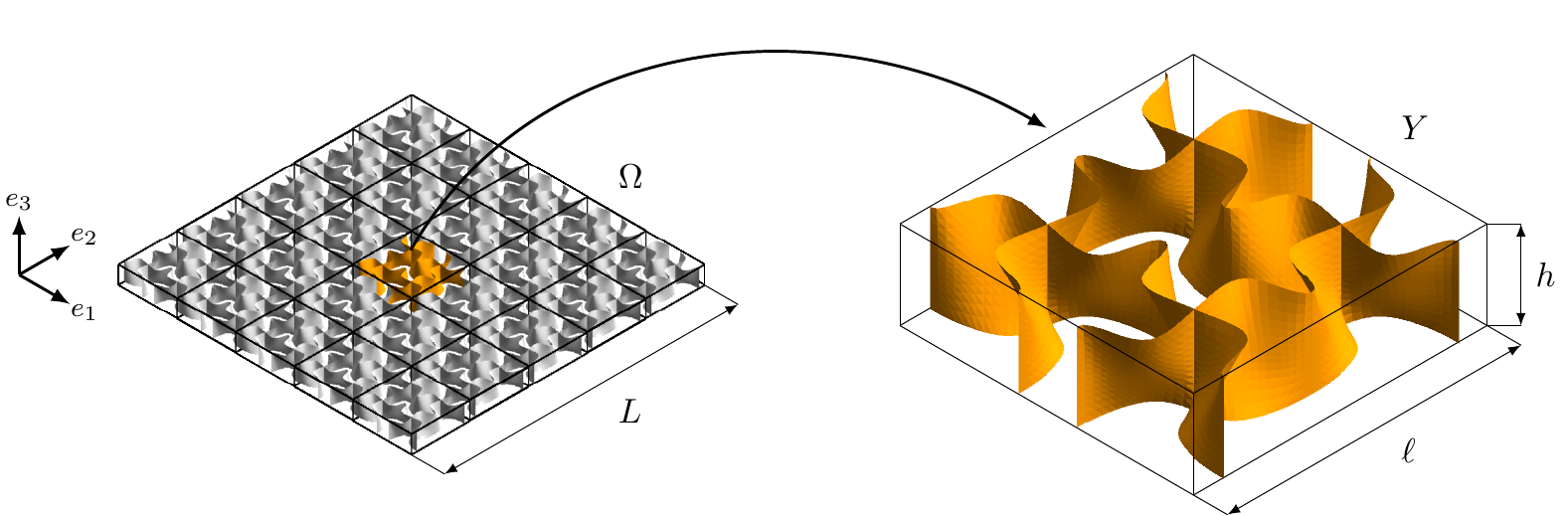}
\caption{Architectured panel $\Omega$ with a periodic arrangement of $5 \times 5$ unit cells. The unit cells are composed of structural shells parametrised by b-spline surfaces. The thickness of the shell $t$ is not depicted (only the mid-surface is shown). The displayed unit cell is obtained following the procedure described in \cref{sec:unit-cell}. It has an aspect ratio $h^* = h /\ell = 0.3$.}
\label{fig:panel-design}
\end{figure}

\section{Design of the unit cell}
\label{sec:unit-cell}
The architectured panel under consideration is a plane composite occupying a domain $\Omega$, of characteristic in-plane dimension $L$, and of height $h$, as sketched in \Cref{fig:panel-design}. It is composed of a large number of quadrangular building blocks, referred to as \textit{unit cells} and denoted as $Y$, of characteristic in-plane dimension $\ell$. They are arranged in a periodic or graded pattern within $\Omega$. Here, we restrict ourself to hollow composites unit cells, similar to open cell foams \cite{Gibson1982a, Ashby2005}, but precisely characterized by a unique ribbon- and membrane-like micro-architectures of constant thickness $t$, as illustrated in \Cref{fig:panel-design}. We also enforce a design continuity between each opposite boundary, to ensure the  connection within the periodic arrangement. Depending on the application, these unit cells may be scaled to various sizes. To overcome the scale effect, we define a rescaled unit cell $Y^*$ characterized by two dimensionless parameters: the unit cell aspect ratio, $h^* = h/\ell$ and the normalized thickness, $t^* = t/\ell$.\smallskip

\begin{figure}
\centering
\includegraphics{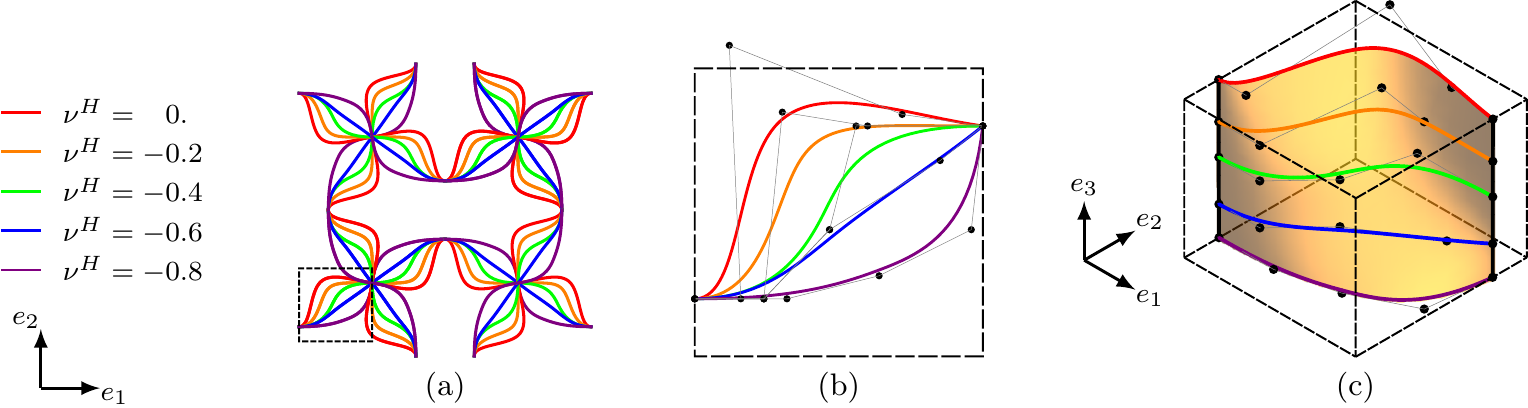}\\[6pt]
\subf{\begin{footnotesize}
\begin{tabular}{|l|lllll|}
\hline
 & CP1 & CP2 & CP3 & CP4 & CP5 \\
\hline
$\nu^H=0$    & $(0.25, 0.25, 0)$        & $(0.18, 0.26, 0)$      & $(0.03,0.32, 0)$
	         & $(0.04, 0.1, 0)$         & $(0, 0.1, 0)$ \\
\hline
$\nu^H=-0.2$ & $(0.25, 0.25, 0.25h^*)$    & $(0.15,0.25, 0.25h^*)$ & $(0.08,0.26, 0.25h^*)$
             & $(0.06, 0.1, 0.25h^*)$     & $(0, 0.1, 0.25h^*)$ \\
\hline
$\nu^H=-0.4$ & $(0.25, 0.25, 0.5 h^*)$    & $(0.14,0.25, 0.5 h^*)$ & $(0.12,0.16, 0.5 h^*)$
             & $(0.06, 0.1, 0.5 h^*)$     & $(0, 0.1, 0.5 h^*)$ \\
\hline
$\nu^H=-0.6$ & $(0.25, 0.25, 0.75h^*)$    & $(0.21,0.22, 0.75h^*)$ & $(0.12, 0.16, 0.75h^*)$ 
             & $(0.06, 0.1, 0.75h^*)$     & $(0, 0.1, 0.75h^*)$ \\
\hline
$\nu^H=-0.8$ & $(0.25, 0.25, h^*)$        & $(0.24, 0.16, h^*)$     & $(0.16, 0.12, h^*)$
             & $(0.08, 0.1, h^*)$         & $(0, 0.1, h^*)$ \\
\hline
\end{tabular}
\end{footnotesize}}{(d)}
\caption{B-spline parametrisation. (a) Class of architected materials, inspired by Clausen et al. \protect \cite{Clausen2015}. The boxed branches on the bottom left are the most basic pattern required to reconstruct the whole unit cell. (b) elementary pattern parametrised using b-spline and their control points. (c) B-spline surface built upon the uniform distribution of the control points in (b) along the thickness. (d) Coordinates of the control points defining each of the five b-spline curves needed to build the surface. The value report regards a cell of characteristic length $\ell = 1$.}
\label{fig:uc_b-spline}
\end{figure}

The material distribution inside the cell is parametrised using a multiple b-spline surfaces. We remind that a b-spline surface of degree $p$ in the $u$ direction and degree $q$ in the $v$ direction, is a bivariate vector-valued piecewise rational function of the form:
\begin{equation}
\textbf{S}(u,v) = \sum_{i=0}^{n} \sum_{j=0}^{m} N_{i,p}(u) \, N_{j,q}(v) \, \textbf{P}_{i,j}
\end{equation} 
where $\textbf{P}_{i,j}$ is a bidirectional control net, \textit{i.e.} list of control points (CP), while $N_{i,p}(u)$ and $N_{j,q}(v)$ are the particular functions called b-spline basis functions \cite{Piegl1997}. The choice of the control points $\textbf{P}_{i,j}$ is made with the aim to engender the extension bending (EBC) effect in the panel. It is realised manually, according to the following design procedure:

\begin{enumerate}[leftmargin =*]
\item We select a particular class of five two-dimensional unit cells based on filaments with equally dispersed negative effective Poisson's ratios within a range of $[-0.8,0.]$, depicted in \Cref{fig:uc_b-spline}(a). Thanks to the multiple symmetries, the parametrisation is restricted to a bundle of elementary corrugated branches, reported in \autoref{fig:uc_b-spline}(b). We model the branches with b-spline curves \cite{Piegl1997}, controlled by five points each. Since all unit cells are sharing common two outermost control points, the amount of independent design parameters is reduced to three points out of five, \textit{i.e.} six distinct parameters — $(x,y)$ coordinates — per curve. In other words, the position of three control points is sufficient to control the apparent Poisson's ratio of each micro-architectured material.

\item The control points of each b-spline curve (from \Cref{fig:uc_b-spline}(b)) are uniformly  distributed along the thickness according to the desired height. Using the b-spline surface parametrization, we construct a \textit{lofted} surface, depicted in \Cref{fig:uc_b-spline}(c). The obtained surface features a two-fold undulation: (1) an in-plane corrugation stemming from the 2-d microstructures and (2) a continuously varying profile along the height. Fixing the outermost control points for all curves ensures a resulting vertical borders, continuity of the shell and facilitates the construction of the final unit cell (displayed in \Cref{fig:panel-design}). Using geometric transformations such as rotations and reflections, we form a lattice of undulated \textit{ribbons} connected to each other periodically. This, in turn, gives a three-dimensional unit cell that verifies periodicity.
\end{enumerate}

We point out that this ribbon-based unit cell encompasses a continuous stacking of two-dimensional shapes with varying effective Poisson's ratio ranging between $-0.8$ and $0$ along the height. This feature permits to recover two-dimensional unit cells that attain any effective Poisson's ratio between $-0.8$ and $0$ by taking a slice of the micro-architectured panel at the corresponding height. This class of unit cells based on filaments with constant thickness was reported in \cite{Clausen2015} and stems from a topology optimization with the objective to exhibit a prescribed effective Poisson's ratio over finite deformations of up to 20\%. Our choice for this basic filament design was motivated by two reasons. First, in comparison with several other shapes in the literature also designed using topology optimization \cite{Gao2018, Zhang2019, Agnelli2020}, these unit cells are featuring high geometrical simplicity (the unit cell are arrangement of curved beams with constant thickness $t$), which is suitable for design flexibility, and manufacturability. Second, they share a common generic configuration that enables a global parametrization using design points and b-spline curves (as shown in Figure 4 in \cite{Clausen2015}) which be exploited next. However, the present design procedure is not specific to this class of unit-cells and can be based on any other shapes which permit a similar parametrisation.\smallskip

Other advantages of a ribbon-based unit cell include both structural and technical aspects: as demonstrated by in several studies of the literature \cite{Wang2014a, Clausen2015, Liu2018}, using curved beams with constant thickness as structural components for two-dimensional structures tends to maintain the auxetic effect at finite strains. We expect this principle to be generalized to three-dimensional panels composed of undulated structural shells. Moreover, ribbons and membranes have a continuous shape, adapted for particular additive manufacturing technologies based for example on wire deposition, such as Fused Filament Fabrication for polymers, or Wire Arc Additive Manufacturing for metals.

\section{Extension bending coupling (EBC) mechanism}
\label{sec:simulations-method}
In the sequel, we investigate the extension-bending coupling (EBC) mechanism at two-scale: first by analysing the behaviour of a single undulated ribbon under tension and bending (see \cref{sec:undulated-ribbon-study}), second by identifying numerically the homogenized behaviour of the panel at small and finite strain (see \cref{sec:panel-homogenization}). The investigation lead to a parameter analysis, where the the mechanical properties are assessed against the geometrical parameters (see \cref{sec:parameter-analysis}).\smallskip

All numerical simulations are conducted using the finite element solver Cast3M 2018 ({\tt www-cast3m.cea.fr}). The parametric b-spline surfaces are triangulated to generate a discrete shell model. The computations are performed using discrete Kirchhoff triangular (DKT) shell elements \cite{Bathe1983, Talaslidis1992}. Regarding the base material modelling, we consider a quasi-incompressible, isotropic, elastic behaviour, with elastic parameters $E^0=0.7599 \, \mathrm{MPa}$ and $\nu^0=0.49$, yielding a \textit{normalized} in-plane elastic stiffness in tension, \textit{i.e.} $A_{1111}^0 = A_{2222}^0 = 1.0 \, \mathrm{MPa}$ under plane stress assumption. In this manner, we get past the influence of the base material stiffness, and focus on the effect of the deformation mechanisms (that involve rotations). Assuming an elastic behaviour, the stiffness components are proportional to the Young's modulus, so values for real applications can be rapidly derived.\smallskip

For convenience, the middle surface of the panel lies in the $(O,e_1,e_2)$ plane, as shown in \Cref{fig:panel-design}. The elementary ribbons that are composing the panel are therefore oriented along directions $e_1$ or $e_2$. From now on, displacements along $e_1$ or $e_2$ will be referred to as in-plane displacements and denoted $u_1$ and $u_2$, while displacements along $e_3$ will be referred to as out of plane displacement and denoted $u_3$. Rotations in the cross sections about $e_3$ will be referred to as in-plane rotations and denoted $r_3$, while rotations about $e_1$ or $e_2$ will be referred to as out of plane rotations and denoted $r_1$ and $r_2$. 

\subsection{Deformation mechanisms of undulated ribbons}
\label{sec:undulated-ribbon-study}
\begin{figure}[ht]
\centering
\includegraphics[scale=1]{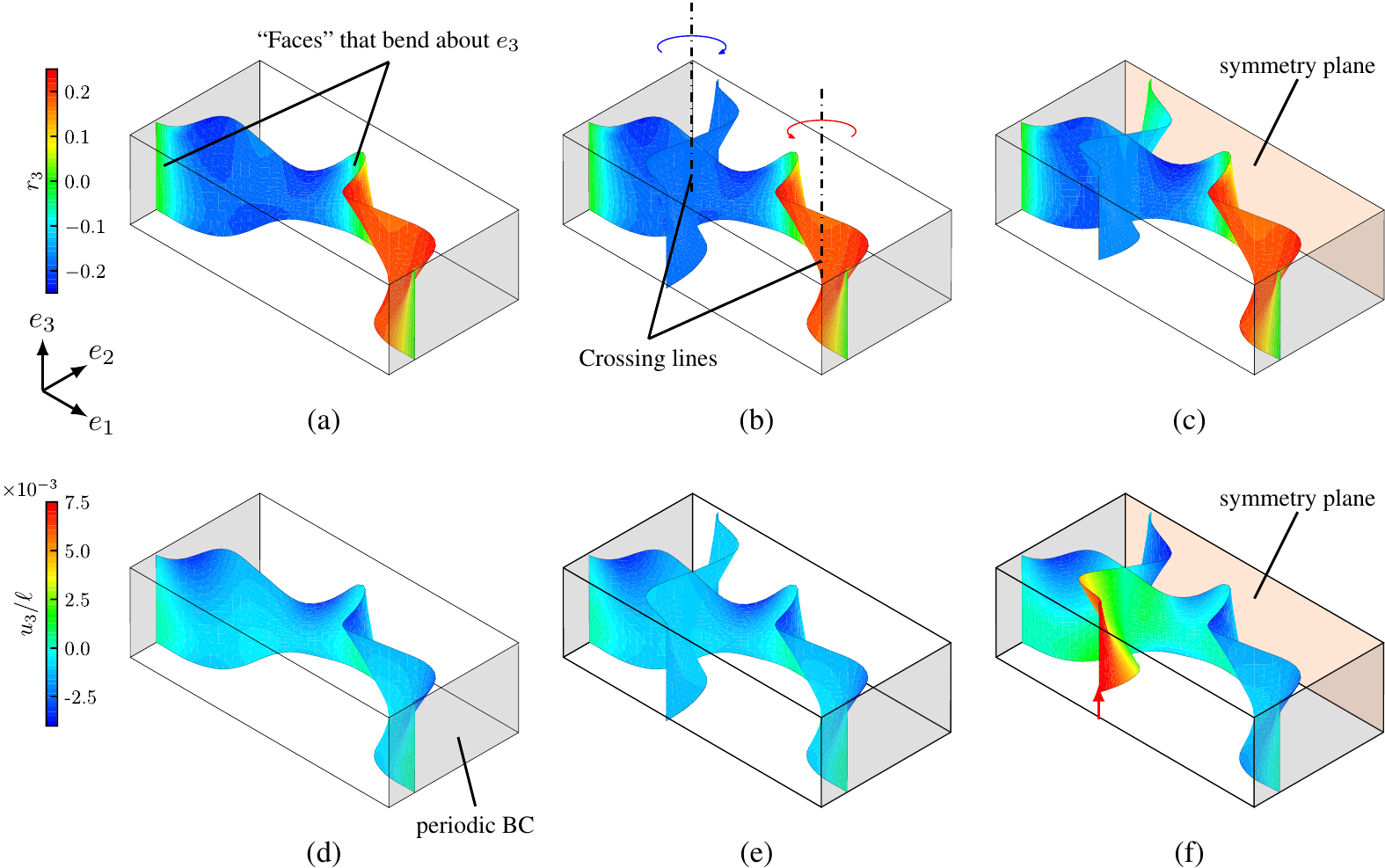}
\caption{Numerical results on ribbon blocks subjected to periodic boundary conditions, loaded in tension up to $10\%$ effective strain. The aspect ratio here is $h^* = h/\ell = 0.3$ and the normalized thickness is $t^* = t/\ell = 0.05$. In all figures, the deformed along the direction $e_3$ was amplified ten times. (a-c) In-plane rotation field $r_3$ (about $e_3$) plotted as a color map on the deformed ribbon. The color bar is the same for the three cases. (d-f) Out of plane displacement field $u_3$ (normalized by $\ell$) plotted as a color map on the deformed ribbon. The color bar on the left applies to the three cases. (a,d) Uniaxial response of a single undulated ribbon. (b,e) Are the same as (a,d), but we attach a transverse undulated ribbon. The free transverse ribbon undergoes a in plane rotation about the $e_3$ axis, but no significant out of the plane displacement is reported. (c,f) The transverse ribbon is submitted to a symmetry condition ($u_2=0$, $r_3=0$), which triggers an out of plane displacement $u_3$.}
\label{fig:ribbon-numerics}
\end{figure}
We begin our investigation by analysing the kinematic deformations modes of elementary undulated ribbons oriented along $e_1$ under a uniaxial tensile load up to $10\%$ effective strain. In particular, we aim to understand the effect of undulated pattern on the deformation mechanisms at the crossing lines, \textit{i.e.} lines of connection with transverse ribbons in the panel that triggers the extension-bending coupling (EBC) effect. We model elementary undulated ribbons of one period, and we impose periodic boundary conditions at the two edges of the ribbon, to ensure the results are not affected by boundary effects.\smallskip

We first consider a single elementary ribbon under tension, whose rotations and displacement fields along $e_3$ obtained numerically are reported in \Cref{fig:ribbon-numerics}(a,d). As the ribbon elongates, the initial in-plane corrugations \textit{unfold} through a bending mechanism about $e_3$ to align with respect to the loading direction $e_1$. The bending mechanisms is localized in particular zones where the cross-section profiles are straight lines oriented along $e_3$ (the zones referred as to ``faces'' that bend about $e_3$ in \Cref{fig:ribbon-numerics}(a)), facilitating the in-plane rotation. At the crossing lines, the displacement remains in-plane and constant along the height ($u_3=0$ as illustrated in \Cref{fig:ribbon-numerics}(d)). Furthermore the out of plane rotations $r_1$ and $r_2$ are vanishing ($r_3$ is constant as illustrated in \Cref{fig:ribbon-numerics}(a)), \text{i.e.} the normal to the ribbons surface at the crossing lines remain in the $(e_1,e_2)$ plane. In \Cref{fig:ribbon-numerics}(b,e), we further verify that even by adding a free transverse ribbon at one crossing line, the system will remain in-plane. As expected, the transverse ribbon deforms in the $(e_1,e_2)$ plane, merely following the rotation $r_3$ dictated by the longitudinal ribbon (as shown in \Cref{fig:ribbon-numerics}(b), the rotation $r_3$ is constant all over the transverse ribbon). All things considered, the deformation mechanism of the two-fold undulated ribbons are no different than a classical folded ribbon. In other words, the specific varying features along the height do not directly induce the EBC effect on single ribbons alone.\smallskip

Next, we consider the same previous set of two ribbons under tension, but we enforce symmetry conditions at the edge of the transverse ribbon, \textit{i.e.} $u_2=0$ and $r_3=0$, as depicted in \Cref{fig:ribbon-numerics}(c,f). Under these conditions, upon pulling on the longitudinal ribbon, the transverse ribbons is submitted to a flexural load about $e_3$. As the longitudinal ribbon elongates, we report a bending of the transverse ribbon, decomposed about $e_3$ (in-plane) and $e_1$ (out of plane). As a result, the longitudinal ribbon is tilted at connecting line, yielding out of plane deflection in the transverse ribbon (\Cref{fig:ribbon-numerics}(f)). We conclude that the shifting mechanism is driven by the coupled in-plane and out of plane flexure that occur inside the ribbons.

\subsection{From undulated ribbons to architectured unit cells}
\label{sec:panel-homogenization}
Next, we analyse the mechanical behaviour of the effective material, \textit{i.e.} the averaged stiffness tensor over the unit cells. For convenience, we introduce the superscripts $H$ when a value is defined at the macroscopic scale. For instance, the effective Young's modulus, Poisson's ratio and shear modulus are denoted as $E^H$, $\nu^H$, and $G^H$ respectively, hereinafter. These three moduli suffice to entirely characterize the in-plane effective behaviour of the architectured material, which exhibits an in-plane \textit{quadratic} symmetry.\smallskip

Due to the great number of unit cells in $\Omega$, the dimension of the periodic cells $\ell$ is assumed to be much smaller than $L$ (\textit{i.e.} $\ell / \epsilon = \mathcal{O}(L)$, where $\epsilon$ tends to 0), but is assumed to be comparable to $h$ (\textit{i.e.} $\ell = \mathcal{O}(h)$). Furthermore, the thickness $t$ is assumed to be much smaller than $\ell$ and $h$ (so that we verify the shell assumption). In practice, we assume:
\begin{equation}
0.1 \leq h^* \leq 10, \qquad t^* \leq 5h^*.
\label{eq:geometry-assumption}
\end{equation}
The effective material properties of the unit cell are evaluated using two-scale homogenization method applied to periodic plates. To interpret the observed bending in terms of effective material parameters, we need to map the behaviour within the classical Kirchhoff-Love plate theory (see \ref{appendix_plate} for a short recall). The homogenisation of plates with periodic microstructure was first studied in \cite{Caillerie1984} and \cite{Kohn1984}. A short recall on the derivation of the linearized effective equations for infinitesimal deformation of panel with periodic microstructure is provided in the \ref{appendix:B}, while interested readers may refer to \cite{Sanchez-Palencia1980, Mei2010} for more extended explanations.\smallskip

The constitutive behaviour of a general Kirchhoff-Love thin plate reads:
\begin{equation}
    \begin{bmatrix} \vec{N} \\ \vec{M} \end{bmatrix} =
    \begin{bmatrix}
    \vec{A} & \vec{B} \\
    \vec{B} & \vec{D}
    \end{bmatrix}
    \begin{bmatrix} \boldsymbol{\mu} \\ \boldsymbol{\chi} \end{bmatrix}
\end{equation}
where:
\begin{itemize}[leftmargin =*]
\item The generalized stresses are the membrane stress $\vec{N}$ and bending moments $\vec{M}$ for unit width. Units are: $[\vec{N}] = \mathrm{N.m^{-1}}$ and $[\vec{M}] = \mathrm{N}$.
\item The plate kinematic is described by the in-plane (membrane) strains $\boldsymbol{\mu}$ and the out-of-plane curvatures $\boldsymbol{\chi}$. Units are: $[\boldsymbol{\mu}] = \mathrm{m.m^{-1}}$ and $[\boldsymbol{\chi}] = \mathrm{m^{-1}}$.
\item The tensor $\vec{A}$ describes the in-plane behaviour, the tensor $\vec{D}$ describes the bending behaviour, and their coupling is expressed through the tensor $\vec{B}$. Units are: $[\vec{A}] = \mathrm{N.m^{-1}}$,  $[\vec{B}] = \mathrm{N}$ and $[\vec{D}] = \mathrm{N.m}$. Note that in most engineering applications, where panels feature symmetric geometry and material distribution along the thickness, normal and shear behaviour get uncoupled for the membrane part, yielding $\vec{B} = 0$.
\end{itemize}
Assuming a composite panel made of two isotropic phases (material and void in this case), the constitutive behaviour for Kirchhoff-Love thin plate exhibits an orthotropic behaviour in the most general case \cite{Sanchez-Palencia1980}, hence it reads in its component form:
\begin{equation}
 \begin{bmatrix}
    \vec{A}^H & \vec{B}^H \\
    \vec{B}^H & \vec{D}^H
    \end{bmatrix}_{(h^*,\, t^*)} =
\begin{bmatrix}
\begin{array}{ccc|ccc}
 A_{1111}^H & A_{1122}^H & 0 & B_{1111}^H & B_{1122}^H & 0 \\[3pt]
 A_{1122}^H & A_{2222}^H & 0 & B_{1122}^H & B_{2222}^H & 0 \\[3pt]
 0 & 0 & A_{1212}^H & 0  & 0 & B_{1212}^H \\[3pt]
 \hline
 B_{1111}^H & B_{1122}^H & 0 & D_{1111}^H & D_{1122}^H & 0 \\[3pt]
 B_{1122}^H & B_{2222}^H & 0 & D_{1122}^H & D_{2222}^H & 0 \\[3pt]
 0 & 0 & B_{1212}^H & 0 & 0 & D_{1212}^H 
\end{array}
\end{bmatrix}
\end{equation}
The in-plane elastic moduli depend on $h$ mainly according to $1/h$, the flexural moduli depend on $h$ mainly according to $1/h^3$. As a matter of fact, $h$ is a parameter which is tending to zero, the plate is then thinner and thinner and it must be more and more rigid to be able to stand the stresses which are applied to it.

\paragraph{Example of effective plate elastic stiffness tensor} In a rescaled unit cell with $h^* = 0.3$ and $t^* = 0.05$, the material volume fraction is of $26.7\%$ and its constitutive tensor reads:
\begin{equation}
 \begin{bmatrix}
    \vec{A}^H & \vec{B}^H \\
    \vec{B}^H & \vec{D}^H
    \end{bmatrix}_{(0.3,0.05)} =
10^{-3}
\begin{bmatrix}
\begin{array}{ccc|ccc}
\phantom{-} 4.18 & -1.89 & 0. &  
\phantom{-} 0.01 & \phantom{-} 0.45 & 0. \\
-1.89 & \phantom{-} 4.18 & 0. &
\phantom{-} 0.45 & \phantom{-} 0.01 & 0. \\
  0.   &  0.   & 0.74 &  0.   &  0.   & 0.44 \\
\hline
\phantom{-} 0.01 & \phantom{-} 0.45 & 0. &
\phantom{-} 1.05 & -0.14 & 0. \\
\phantom{-} 0.45 & \phantom{-} 0.01 & 0. &
-0.14 & \phantom{-} 1.05 & 0. \\
  0.   & 0.    & 0.44 &  0.   &  0.   & 1.02 \\
\end{array}
\end{bmatrix}
\end{equation}
The matrices $\vec{A}^H$, $\vec{B}^H$ and $\vec{D}^H$ are symmetric and exhibits an ``quadratic'' symmetry, \textit{i.e.} the plate has the same tensile (respectively bending) stiffness along $e_1$ and $e_2$. $A_{1122}^H$ is negative, which indicates that the unit cell display an effective auxetic behaviour. An underlying effect of the auxetic behaviour is to undergo synclastic curvatures \cite{Lakes1987}, \textit{i.e.} shifting from flat to a dome shape, which is accounted in the negative $D_{1122}^H$. $B_{1122}^H$, which links the in-plane stress along $e_1$ (respectively $e_2$) to the transverse curvature about $e_1$ (respectively $e_2$) and vice-versa, is the main non-zero coefficients of $\vec{B}^H$. This suggests that the coupling effect exists primarily between the longitudinal in plane displacements and the out of plane bending curvature, which is in agreement with the concept of overlaying profiles with varying Poisson's ratios. We also note the presence of a coupling between the shears $B_{1212}^H$. The elastic moduli, $A_{\alpha\beta\gamma\delta}^H$, $B_{\alpha\beta\gamma\delta}^H$ and $D_{\alpha\beta\gamma\delta}^H$ can be expressed in terms of materials parameters $E^H$ and $\nu^H$. A simple calculation immediately yields:
\begin{equation}
\begin{aligned}
\displaystyle E^H = \frac{1}{h^*} A^H_{1111}\left( 1 - \left(\frac{A^H_{1122}}{A^H_{1111}}\right)^2 \right) = \frac{1}{h^*} A^H_{2222}\left( 1 - \left(\frac{A^H_{1122}}{A^H_{2222}}\right)^2 \right) \, ;& \quad &
\displaystyle \nu^H = \frac{A^H_{1122}}{A^H_{1111}} = \frac{A^H_{1122}}{A^H_{2222}};
\end{aligned}
\end{equation}
We further introduce the effective longitudinal extension - transverse bending coupling ratio, denoted by $\beta^H$. This coupling ratio is defined in the same spirit than the Poisson's ratio:
\begin{equation}
\beta^H = \frac{1}{h} \frac{B^H_{1122}}{A^H_{1111}} 
        = \frac{1}{h} \frac{B^H_{1122}}{A^H_{2222}}
\label{eq:beta}
\end{equation}
\paragraph{Note} the division by $h$ is here to obtain a dimensionless quantify.

\begin{figure}
\centering
\subf{\includegraphics[width=0.48\textwidth]{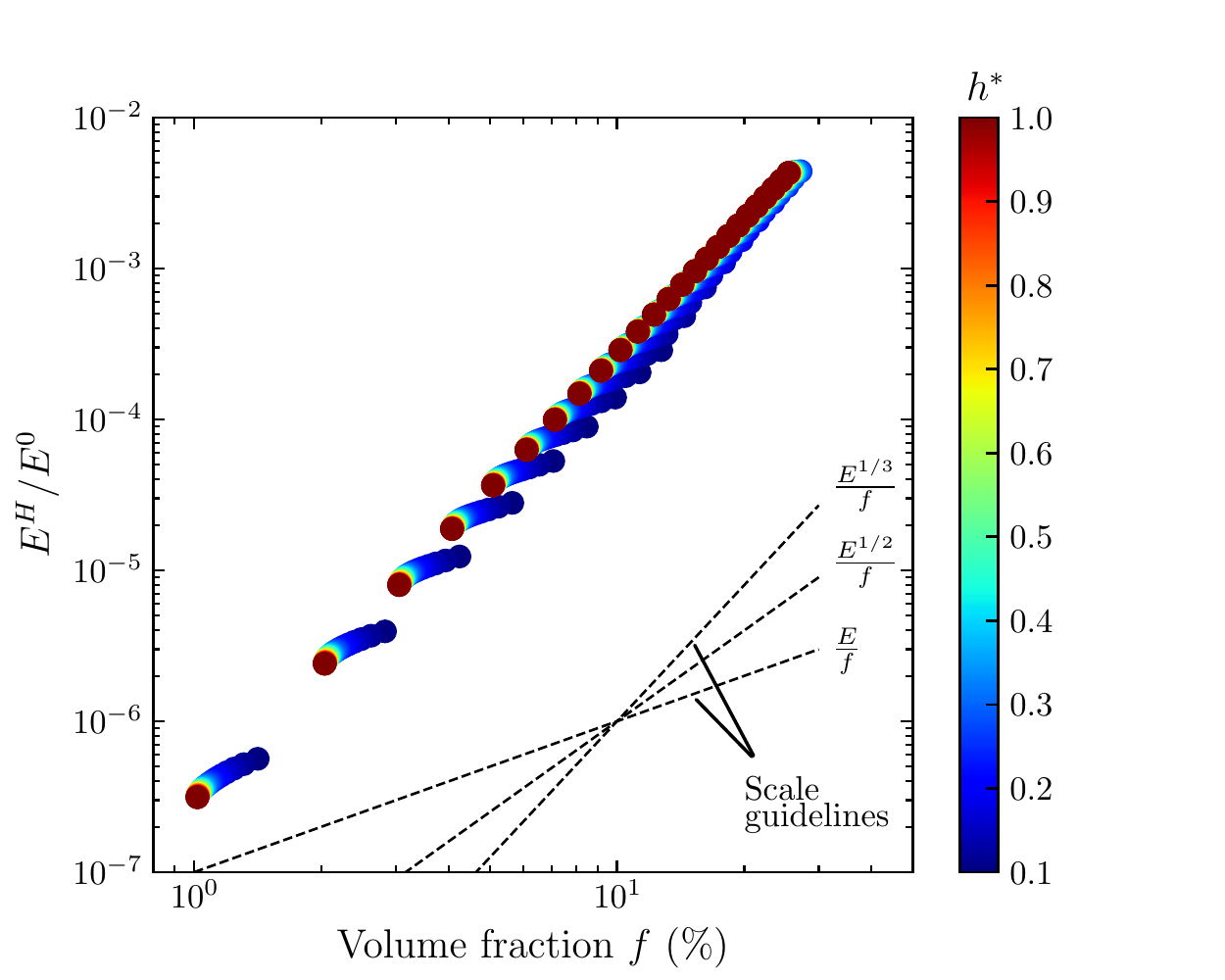}}{(a)}
\hfill
\subf{\includegraphics[width=0.48\textwidth]{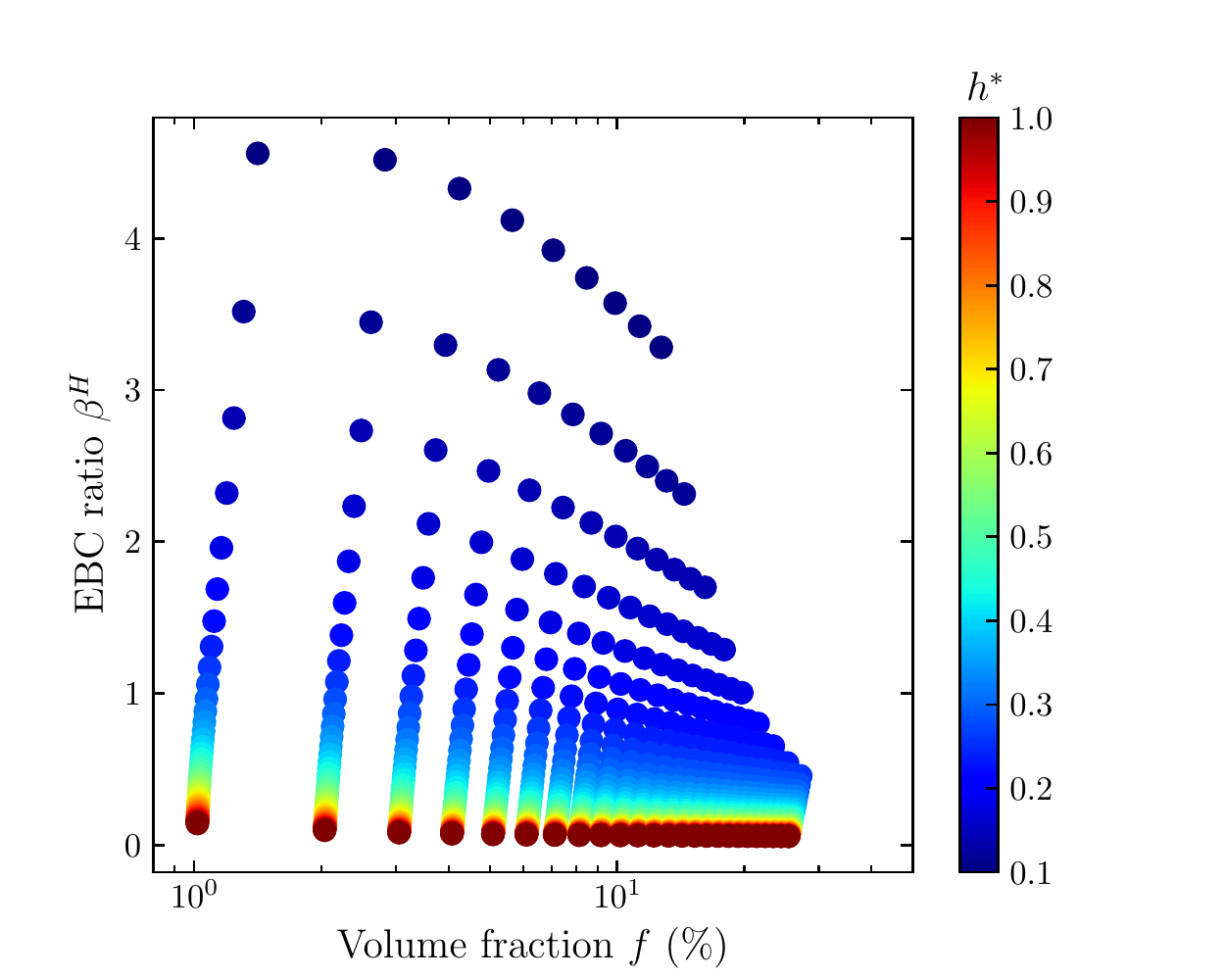}}{(b)}
\caption{Parameter analysis with respect to $h^*$ and $t^*$. Material property charts (a) Normalized effective Young's modulus $E^H/E^0$ versus material volume fraction $f$. Note that the performances of these architectured materials can be mapped in already existing material property charts, like in \protect \cite{Fleck2010}, for further comparisons with other materials. (b) EBC ratio $\beta^H$ from equation \eqref{eq:beta} versus material volume fraction.}
\label{fig:small-strain-properties}
\end{figure}

\subsection{Influence of \texorpdfstring{$h^*$ and $t^*$}{Lg}}
\label{sec:parameter-analysis}
Having demonstrated that unit cells based upon undulated ribbons can exhibit EBC effect, we now investigate the effect on the mechanical properties of such structure, of the unit cell aspect ratio of the panel $h^*$, the ribbons normalized thickness $t^*$. The parameter analysis is performed by varying: (1) $h^*$ from $0.1$ to $1$ with a step of $0.02$ and; (2) $t^*$ from $0.002$ to $0.05$ with a step of $0.002$. We point out however that cases where assumptions on $h^*$ and $t^*$ from equation \eqref{eq:geometry-assumption} do not apply are excluded. The results are reported in against the material volume fraction $f$ in the cell, to facilitate possible comparisons with other materials property charts \cite{Fleck2010}. Volume fraction, estimated from the three-dimensional numerical models, is depending linearly to $t^*$, while it is almost unaltered by $h^*$.

\paragraph{Investigation under small strain assumption} The section is devoted to studying the range of achievable $E^H$, and EBC ratio $\beta^H$. The distribution of effective Young's modulus $E^H$, normalised by the base material modulus $E^0$, is mapped against the volume fraction $f$ for different values of $h^*$ in \Cref{fig:small-strain-properties}(a). It is proportional to the cube of the volume fraction. Conversely, its dependence on the aspect ratio $h^*$ is much less pronounced, \textit{i.e.} a unit cell of aspect ratio $h^*$ is about as stiff as a pile of $n$ unit cells of aspect ratio $h^*/n$. Macroscopically, this ribbon based unit cell is highly compliant, the Young's modulus $E^H$ being from two to six orders of magnitude lower than its bulk equivalent (see base elastic coefficients in section \ref{sec:simulations-method} above). Next, the distribution of the EBC ratio $\beta^H$ indicates the effect is stronger for smaller aspect ratios $h^*$. Moreover, we report that for $h^* > 0.25$, the EBC ratio $\beta^H$ is almost independent from the normalized thickness $t^*$, while for $h^*<0.2$, the EBC ratio is affected by $t^*$. All things considered, $E^H$ and $\beta^H$ can be tailored relatively independently. On a side note, the Poisson's ratio $\nu^H$ does not depend particularly depend on $h^*$ and $t^*$. Its average value, around $-0.4$ match the expectations considering that the initial two-dimensional profiles had effective Poisson's ratio ranging between $-0.8$ and $0$.

\begin{figure}
\centering
\includegraphics[scale=1]{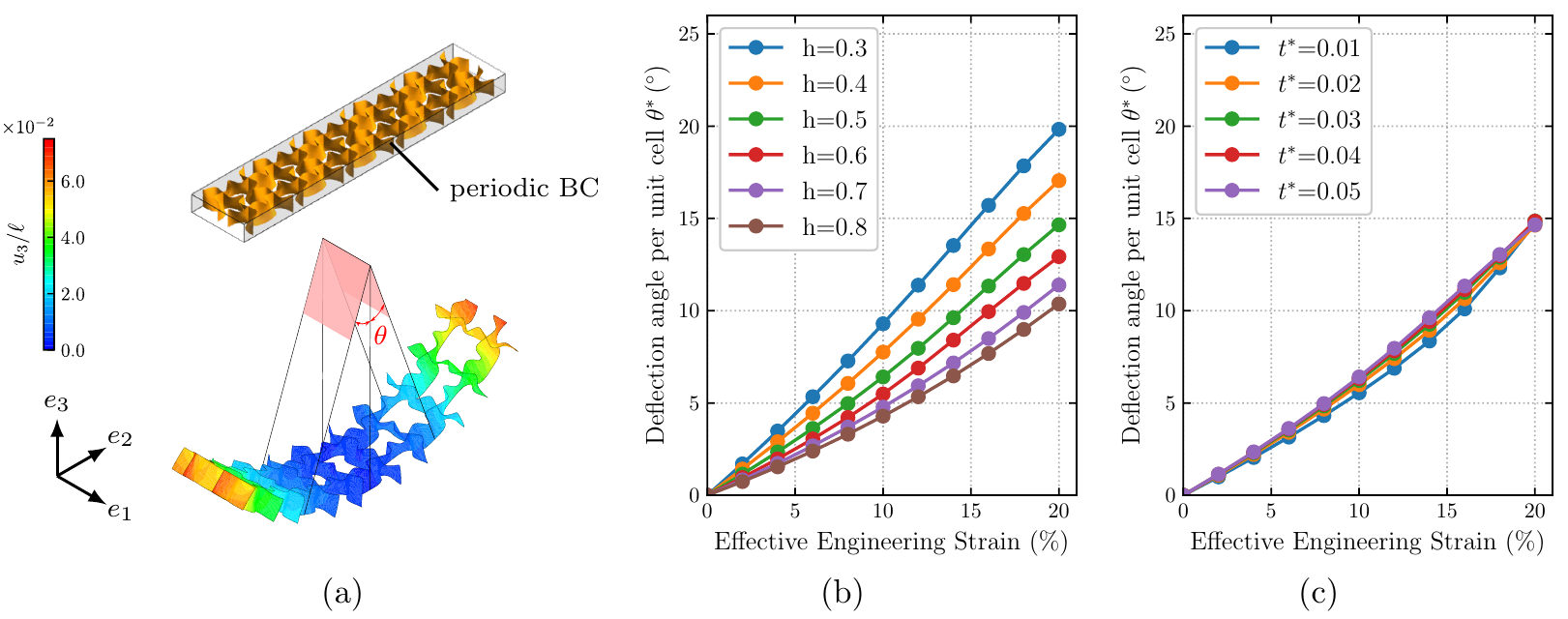}
\caption{(a) Deformed configuration under tension along $e_1$ up to 20\% effective strain, with imposed periodic BC at front and rear faces. The deflection angle $\theta^*$ per unit cell induced in the direction perpendicular to the loading is illustrated. (b) Evolution of $\theta^*$ against the longitudinal effective engineering strain. $t^*$ is set at 0.05, $h^*$ is varying. (c) Same as (b), but now $h^*$ is set at 0.5, $t^*$ is varying.}
\label{fig:curvature_graph}
\end{figure}

\paragraph{Extension at finite strain} We now focus on the out of the plane kinematic capacities at finite strain up to $20\%$, hence we analyse the transverse angle deflection per unit cell $\theta^*$, obtained from a longitudinal tensile loading. A representation of $\theta^*$ is provided in \Cref{fig:curvature_graph}(a). Recall that the two-dimensional profiles designed in \cite{Clausen2015} maintain their behaviour up to $20\%$ effective strain state. To ensure the results are not affected by boundary effects, we analyse the central unit cell in an infinite periodic system. Along the $e_1$ direction, we apply boundary conditions, while along $e_2$ we repeat a large number of cell to enable the panel to bend. Note that only $5$ unit cells are represented in \Cref{fig:curvature_graph}(a), yet our simulation included a symmetric boundary conditions on one side of the panel, which comes down to considering a row of $10$ unit cells. For different loading values, we collect the nodes at the two lateral boundary of the central cell (normal vectors to the faces are $-e_1$ and $e_1$), which form two planes (depicted in pink in \Cref{fig:curvature_graph}(a)). A least square fit permits to calculate the cartesian equation of these two planes, which in turn yields the deflection angle $\theta^*$. The observation made at small strain on the coupling behaviour remain valid at large strain. The EBC mechanism, illustrated through the transverse deflection angle in \Cref{fig:curvature_graph}(b,c), increases for thinner panels (when $h^*$ is small). For $h^*=0.5>0.25$, the it is almost unaffected by $t^*$. We further highlight that for the configurations depicted in \Cref{fig:curvature_graph}(b,c), the evolution of the deflection angle with respect to the engineering strain follows a linear trend. This indicates that the EBC ratio remains constant for deformations ranging up to $20\%$.

\section{Analysis of fabricated polymer panels}
\subsection{Additive manufacturing}
To verify the effective properties, specimen of architectured panels with ribbon-based unit cells have been additively manufactured with fused filament fabrication technology (FFF) using a commercial Ultimaker 3-d printer. We selected a thermoplastic polyurethane (TPU 95A) as the base material because of its suitable compliant nature, and capacity to undergo large deformations ($>20\%$ strain) without breaking. Characteristics of this material can be found in the manufacturers data sheet\footnote{https://support.ultimaker.com/hc/en-us/sections/360003556679}. The material was provided as filament of diameter $0.4\textrm{mm}$, which was fed from a large spool through a moving, heated printer extruder head, and was deposited on the growing work. The print head is moved under computer control to define the printed shape.\smallskip

The 3-d ``bulk'' model of the unit cell was obtained by computing the normal vectors field of the neutral membrane b-spline surface and by creating upper and lower shifted surfaces.  these shifted b-spline surfaces are triangulated to recover a surface mesh. We also mesh the top and bottom boundary, to obtain a closed envelope that can be exported as a watertight STL mesh of the specimen. For the printing, we set $h^*=0.3$ and $t^*=0.05$ and we create a periodic array of $5 \times 5$ unit cells. The characteristic length of the unit cell $\ell$ is set at $16 \, \mathrm{mm}$, thus the dimensions of the whole panel (supports excluded) are $80 \, \mathrm{mm} \times 80 \, \mathrm{mm} \times 4.8 \, \mathrm{mm}$ (the in-plane characteristic dimension here is $L=80 \, \mathrm{mm}$). In order to perform a mechanical test under uniaxial tensile loading, the generated pattern is completed by a series of rings to ensure the fixture to the testing machine. The motivation to employ non-conventional fixture with rings is discussed in the next section.

\subsection{Mechanical testing}

\begin{figure}[ht]
\centering
\begin{tikzpicture}
\node[anchor=west,inner sep=0] at (0,0) {\subf{\includegraphics[height=5cm]{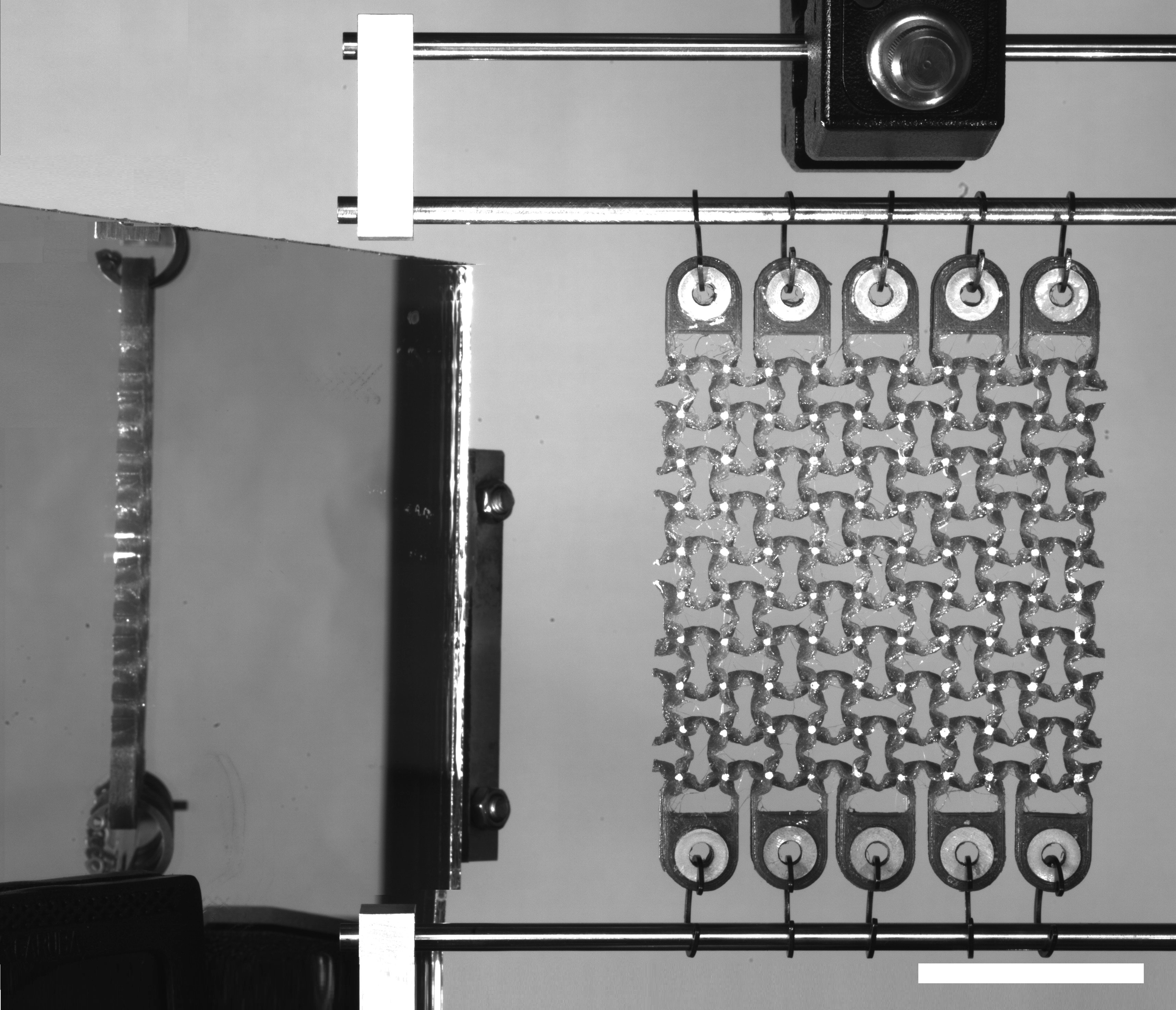}}{(a)}};
\draw[line width=0.3mm] (0.5, 1.7) -- (2,3.3);
\node[right] at (2,3.3) {{\footnotesize \textit{Mirror tilted at $45^{\circ}$}}};
\node[anchor=west,inner sep=0] at (7,0) {\subf{\includegraphics[height=5cm]{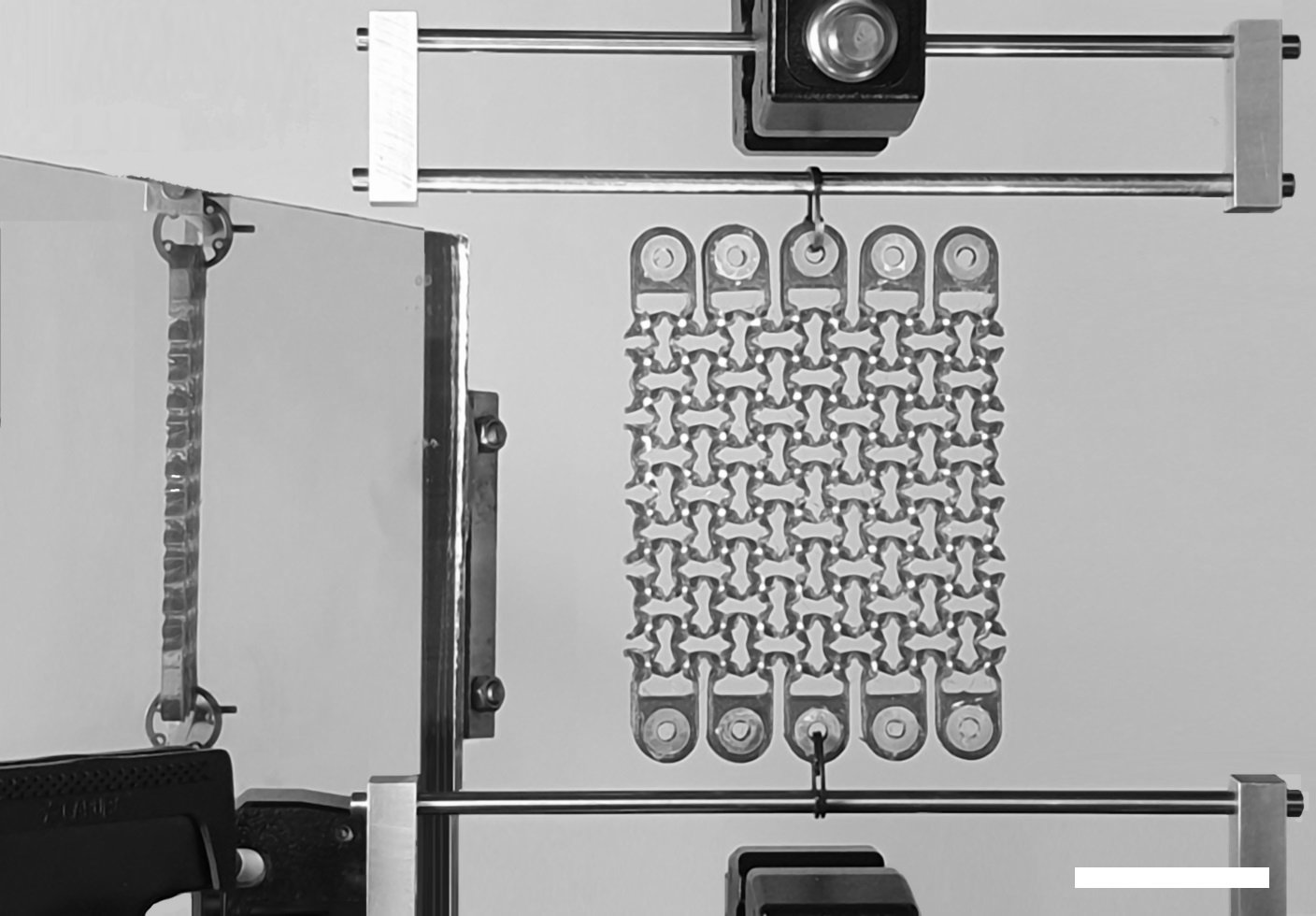}}{(b)}};
\draw[line width=0.3mm] (7.5, 1.8) -- (9,3.3);
\node[right] at (9,3.3) {{\footnotesize \textit{Mirror}}};
\end{tikzpicture}
\caption{(a) Setup for the tensile test. Rings are held together at their edges and are attached to the (sliding) upper grip. Conversely, the central rectangular rod is attached to the (fixed) lower grip. (b) Setup for the tensile test with point-like boundary. Scale bar is 40 mm.}
\label{fig:testing}
\end{figure}
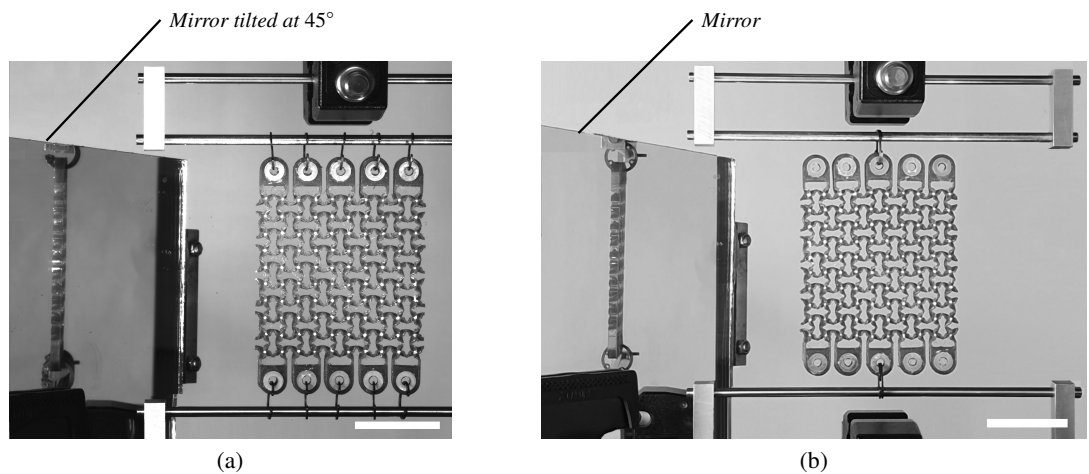

We conclude the characterization of this ribbon based architectured panel through an experimental validation. The experiments are conducted under displacement control at a quasi-static strain rate $\dot{\varepsilon}=0.125 \mathrm{min}^{-1}$ up to $20\%$ effective engineering strain for the tensile test. The tests are performed on an Instron 10 kN universal testing machine. We aim to estimate the out of plane displacement and compare it to our shell-based numerical model.\smallskip

The experiments were piloted using the Instron BlueHill software. Each mechanical test was recorded using a high-resolution digital camera (JAI Spark SP-20000-USB camera with a resolution of $5120 \times 3840$ pixels equipped with a Tokina AT-X Pro 100 mm F2.8 macro lens), mounted on a perpendicular axis with respect to the plane of the specimen. Each picture includes the front and lateral view of the specimen, as shown in \Cref{fig:testing}. To easily achieve optical measurements, a white grid was added to the sample. Using an in-built computer program, 8-bit gray scale sub-images were stored every second during the loading. The sequence of image permits is used for a 2-d point tracking method (undertaken  with the software Tracker, \texttt{https://physlets.org/tracker/}), where we measure the out of plane displacement $u_3$ at the tip of the ribbon identified in the lateral (mirror) view.\smallskip

Early stages tests performed on specimens with standard hard clamp montages (not shown here)  gave unsatisfactory results, \textit{i.e.} the resulting out of plane displacement field was below the expected results of \Cref{fig:curvature_graph}. This is a clear indication that the specimen can not be approximated as an infinite periodic domain in its the centre in this testing configuration assuring: $L \gg \ell$. We deduce that classical clamping induces over-constraining boundary conditions, which prevent the specimen to undergo out of the plane deformation. To bypass this problem, general recommendations include increasing the number of unit cells in the panel such that $L \gg \ell$ is effectively satisfied, and/or ``relaxing'' the clamped  boundary conditions. The solution discussed next makes use of a modified setup to fix the specimen in the tensile machine, which frees the boundary conditions.

\paragraph{Homogenous extension} In this experiment, the specimen is hung to the clamps at both extremities via metallic rings to steel rods in a curtain-like fashion, as shown in \Cref{fig:testing}. Prior to the montage, the steel rods were covered in oil to reduce the frictions with the rings. This fixture accommodates lateral expansions and/or contractions of the specimens undergoing tensile loads, but also rotations (the rings can easily tilt). Similar specimen fixing were attempted in previous works in the literarure \cite{Neville2016, Celli2018, Wang2019}, to decrease the constraints on the boundary conditions ans obtain a homogeneous state of strain. However, it is worth noting that these fixings introduce higher uncertainties on boundary conditions and stress state, due to unknown friction state  between the rings and the axes.\smallskip

Therefore, the numerical computation were undertaken with two distinct types of boundary conditions. On one hand, we considered free traction forces in the respective directions, which correspond to the assumption that friction can be neglected. In practice, it implies that the rings can freely slide and rotate along the supporting axis. As a consequence field will be closer to the homogeneous solution obtained by imposing perfect periodic boundary conditions. On the other hand, we modelled the boundary conditions between the rings and the rods as pivots, which correspond to the assumption that adherence prevents the rings to freely slide. In the simulation, the rings are modelled by thin bars, that cannot slide but can tilt at their extremities. \smallskip

\begin{figure}[ht]
\centering
 \includegraphics{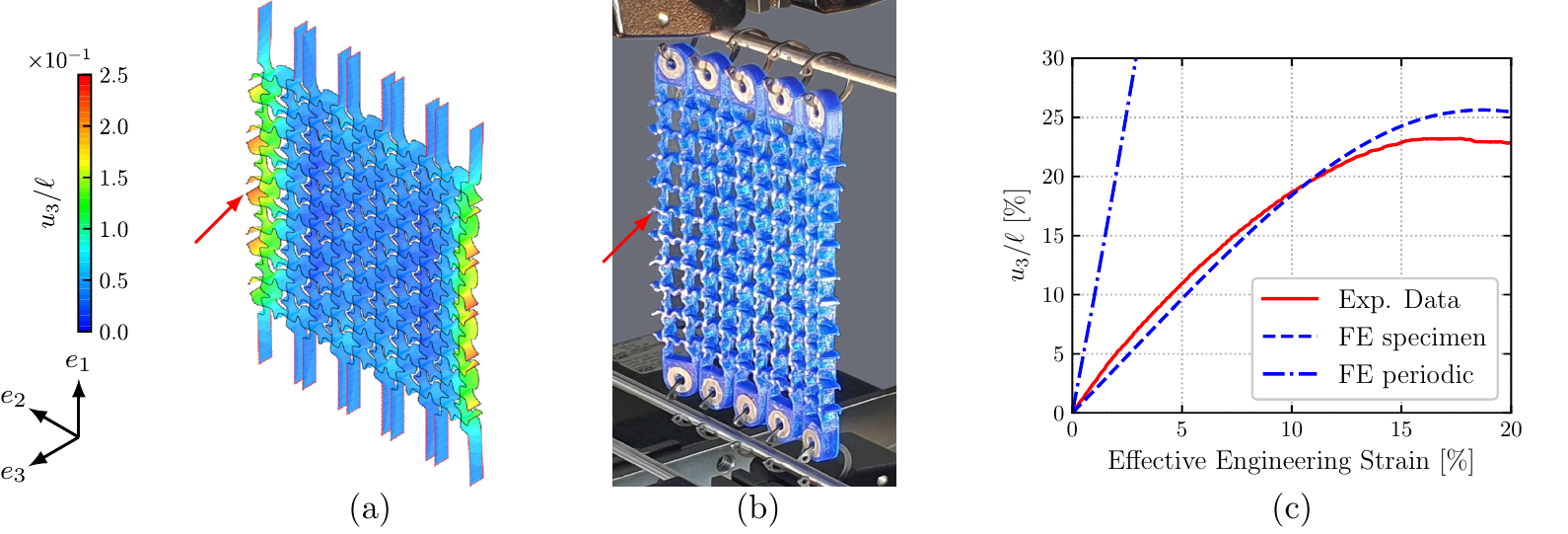}\\[3pt]
\includegraphics{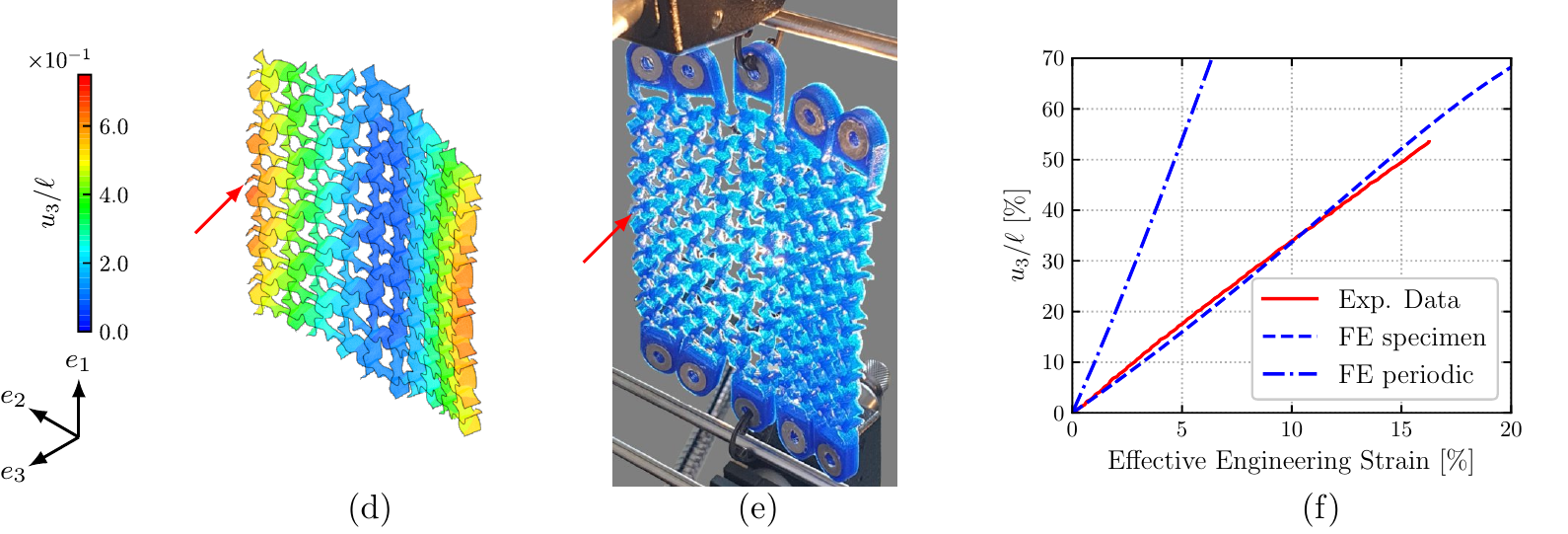}
\caption{(a-c) Tensile test. (a) Numerical and (b) experimental deformed configurations of specimens at imposed engineering effective strain of $20\%$. The out of plane displacement $u_3$ is plotted as a color map. (c) Out of plane displacement. (d-f) Pinching test. (d) Numerical and (e) experimental deformed configurations of specimens at imposed engineering effective strain of $10\%$. The out of plane displacement $u_3$ is plotted as a color map. (f) Out of plane displacement. The red arrows indicate the chosen point to track the out of plane displacement.}
\label{fig:testing-results}
\end{figure}

The results of the extension experiment are represented in \Cref{fig:testing-results}. The different panel represent: (a) the non-dimensionalised computed out of plane displacement field $u_3/\ell$ plotted on the deformed mesh, (b) the observed specimen during the experiment at maximal extension, \textit{i.e.} $20\%$ effective strain and (c) a comparison of the experiment with computations under periodic and experimental boundary conditions of the out of plane displacement of the mid-point of the lateral surface. The computations under experimental boundary conditions match the experimental measurements and exhibit the saturation of the out-of-plane displacement at about $15\%$ engineering strain. Further differences could be explained by the neglected friction and further differences of the boundary conditions,  as experimental observation showed that rings did rotate but did slide during the applied extension. The experiment shows however an important discrepancy with the behaviour of the structure under perfect periodic boundary condition. Under periodic boundary conditions the out-of-plane effect increased by a factor of six and the out-of-plane displacement is proportional with applied extension strain. Additional insight of the deformation pattern can be obtained by a comparison with \Cref{fig:curvature_graph}. 

\paragraph{Extension by concentrated force, i.e. pinching test} In the sequel, this test is referred as to \textit{pinching test}. In order to further relax boundary conditions, only the central ring is loaded in tension. The boundary condition in the numerical computation of the experiment are blocked displacement and the rotations on the two central ribbons. The results of the experiments and computations are resumed as before in \Cref{fig:testing-results}: (d) computed out of plane displacement field $u_3$ normalised by $\ell$ plotted on the deformed mesh, (e) the observed specimen during the experiment at maximal extension, i.e. here $16\%$ effective strain and (f) a comparison of the experiment with computations under periodic and experimental boundary conditions of the out of plane displacement of the mid-point of the lateral surface. The results exhibit a three-fold increase of the out-of-plane displacement and the shape change is easily spotted on the image, see \Cref{fig:testing-results}(e). As before one can remark an excellent match between experiments and computations. A clear enhancement of the out-of-plane displacement can equally be observed when comparing the result with the case of periodic boundary conditions. Indeed, the  periodic case exhibit only a two-fold increase with respect to the pinched specimen. Again, when comparing the pinching experiment with the computation under periodic boundary conditions from \Cref{fig:curvature_graph}, one should notice that only the central cell could eventually reach the desired boundary conditions, as the cells at the lateral edges of the specimen do not experience any vertical force.   

\section{Conclusion and perspectives}
In this paper, we designed a new class of micro-structures composed of undulated ribbon lattice. These micro-structures can be arranged periodically to obtain panels exhibiting a bending deflection when submitted to in-plane tension, hence producing the EBC effect. For a prescribed shape of the of the unit cell, \textit{i.e.} shape of the ribbons and intersections, the aspect ratio of the ribbons and their thickness, tailor the various elastic coefficients, like stiffness or EBC ratio.\smallskip

Our work expands the spectrum of shape-morphing structures manufactured with a single material, and it indicates an approach that could be used to produce morphing and deployable structures for a wide range of scales. While the shapes we have obtained are relatively simple, similar principles could be extended to different families of materials, and could be coupled to parametric optimization (using b-spline) and inverse-design strategies to obtain more extreme shapes. This type of analysis would also permit to shed light on the set of realizable moduli using undulated ribbon-based structures (refer for analogy to the study of \cite{Milton1995} regarding laminates).\smallskip

We foresee these class micro-structures in more advances arrangements where they are repeated, transformed and laid out to generate a pattern with desired morphing capabilities. Breaking the periodicity in the design can be leveraged to achieve more complex shapes, with a local control of the curvature \cite{Celli2018}. The shape-shifting functionalities can also be expanded by controlling the stiffness within the panel, \textit{e.g.} by printing the structures with multiple materials of different stiffness or thermal expansion coefficient. On top of that, we can imagine to embed a self-shaping property in the panel, via the use of responsive materials like shape memory polymers.

\section*{Acknowledgements}
F. A. acknowledges the support of the French doctoral fellowship ``Contrat Doctoral Spécifique pour Normalien''. The specimens have been printed at the Drahi-X Novation Center from Ecole polytechnique during the Summer, 2020. The authors acknowledge Aline Becq and Gareth Paterson for valuable input regarding the printing aspects and Giovanni Frascella, for his help with the testing phase.

\bibliographystyle{elsarticle-num}

\pagebreak
\appendix
\section{Recall on Kirchhoff-Love plate theory}
\label{appendix_plate}
In a space endowed with an orthonormal reference $(O,\vec{e_1},\vec{e_2},\vec{e_3})$, let us consider a plane plate of thickness $h$ normal to the axis $(O, \vec{e_3})$. For convenience, its mid-plane in the reference configuration is assumed to lie in the $(O, \vec{e_1}, \vec{e_2})$ plane.

\paragraph{Thin plate kinematics} The following kinematic assumptions are made: (1) all straight lines normal to the mid-surface remain straight and normal after deformation; (2) the thickness of the plate does not change during a deformation. This is the plate equivalent of the Euler-Bernoulli beam hypothesis. The displacement field $\vec{u}(x_1,x_2,x_3)$ of a thin planar plate is therefore defined as follows:
\begin{equation}
\begin{aligned}
& \phantom{\{}\vec{u}(x_1,x_2,x_3)= \vec{v}(x_1,x_2)- x_3 \, \nabla v_3(x_1,x_2) \\
\Leftrightarrow & 
\begin{cases}
u_1(x_1,x_2,x_3)= v_1(x_1,x_2)-x_3\, v_{3,x_1}(x_1,x_2)= v_1(x_1,x_2)-x_3 \, r_{3}(x_1,x_2) \\
u_2(x_1,x_2,x_3)= v_2(x_1,x_2)-x_3\, v_{3,x_2}(x_1,x_2)= v_2(x_1,x_2)-x_3 \, r_{3}(x_1,x_2) \\
u_3(x_1,x_2,x_3)= v_3(x_1,x_2)
\end{cases}
\end{aligned}
\end{equation}
where $\vec{v}(x_1,x_2)$ is the displacement field of the mid-plane of the plate, and $\vec{r}(x_1,x_2)= \nabla v_{3}(x_1,x_2)$ are the rotations. Assuming the previous displacement field, the strain field $\boldsymbol{\varepsilon}$ reads:
\begin{equation}
\boldsymbol{\varepsilon} = 
\begin{bmatrix}
	\varepsilon_{11} & \varepsilon_{12} \\           
    \varepsilon_{12} & \varepsilon_{22} \\
\end{bmatrix}
=
\begin{bmatrix}
	\displaystyle \frac{\partial v_1}{\partial x_1} & \displaystyle \frac{1}{2}\left(\frac{\partial v_1}{\partial x_2} + \frac{\partial v_2}{\partial x_1}\right)\\           
    \displaystyle \frac{1}{2}\left(\frac{\partial v_1}{\partial x_2} + \frac{\partial v_2}{\partial x_1}\right) & \displaystyle \frac{\partial v_2}{\partial x_2} \\
\end{bmatrix} 
- x_3
\begin{bmatrix}
	\displaystyle \frac{\partial^2 v_3}{\partial x_1^2} & \displaystyle \frac{\partial^2 v_3}{\partial x_1 \partial x_2}\\
     \displaystyle \frac{\partial^2 v_3}{\partial x_1 \partial y_1} & \displaystyle \frac{\partial^2 v_3}{\partial x_2^2} \\
 
\end{bmatrix} =
\boldsymbol{\mu} + z \boldsymbol{\chi}      
\label{compatibilty_plate}
\end{equation}
where $\boldsymbol{\mu}$ represents the in-plane strains and $\boldsymbol{\chi}$ the out-of-plane curvatures. Units are: $[\boldsymbol{\mu}] = \mathrm{m.m^{-1}}$ and $[\boldsymbol{\chi}] = \mathrm{m^{-1}}$. Note that the out-of-plane strains $\varepsilon_{i3}$ are all zero due to the chosen kinematic hypothesis. In particular, the normal out-of-plane strain $\varepsilon_{33}$ is zero, which is generally not the case for thin structures, for which the plane stress behaviour is assumed.

\paragraph{Constitutive behaviour} 
The constitutive law of a thin plate has the following form:

\begin{equation}
    \begin{bmatrix} \vec{N} \\ \vec{M} \end{bmatrix} =
    \begin{bmatrix}
    \vec{A} & \vec{B} \\
    \vec{B} & \vec{D}
    \end{bmatrix}
    \begin{bmatrix} \boldsymbol{\mu} \\ \boldsymbol{\chi} \end{bmatrix}
\end{equation}
where the tensor $\vec{A}$ describes the in-plane behaviour, the tensor $\vec{D}$ describes the bending behaviour, and their coupling is expressed through the tensor $\vec{B}$. The generalized stresses are the membrane stress $\vec{N}$ and bending moments $\vec{M}$, defined as follows:
\begin{equation}
\begin{cases}
   \vec{N} = \displaystyle \int_{-h^*/2}^{h^*/2}     \boldsymbol{\sigma} dx_3 \\[12pt]
   \vec{M} = \displaystyle \int_{-h^*/2}^{h^*/2} x_3 \boldsymbol{\sigma} dx_3
\end{cases}
\end{equation}
where $\boldsymbol{\sigma}$ denotes the stress field.
\pagebreak

\section{Two-scale analysis and effective coefficients}
\label{appendix:B}
Since the panel thickness $h$ is comparable to the unit cell size $\ell$, we only need two-dimensional macroscopic coordinates $\vec{x} = (x_1, x_2)$ for in-plane variations:
\begin{equation}
x_1 = \epsilon y_1, \, x_2 = \epsilon y_2
\end{equation}

\begin{figure}
\centering
\subf{\includegraphics[width=0.25\columnwidth]{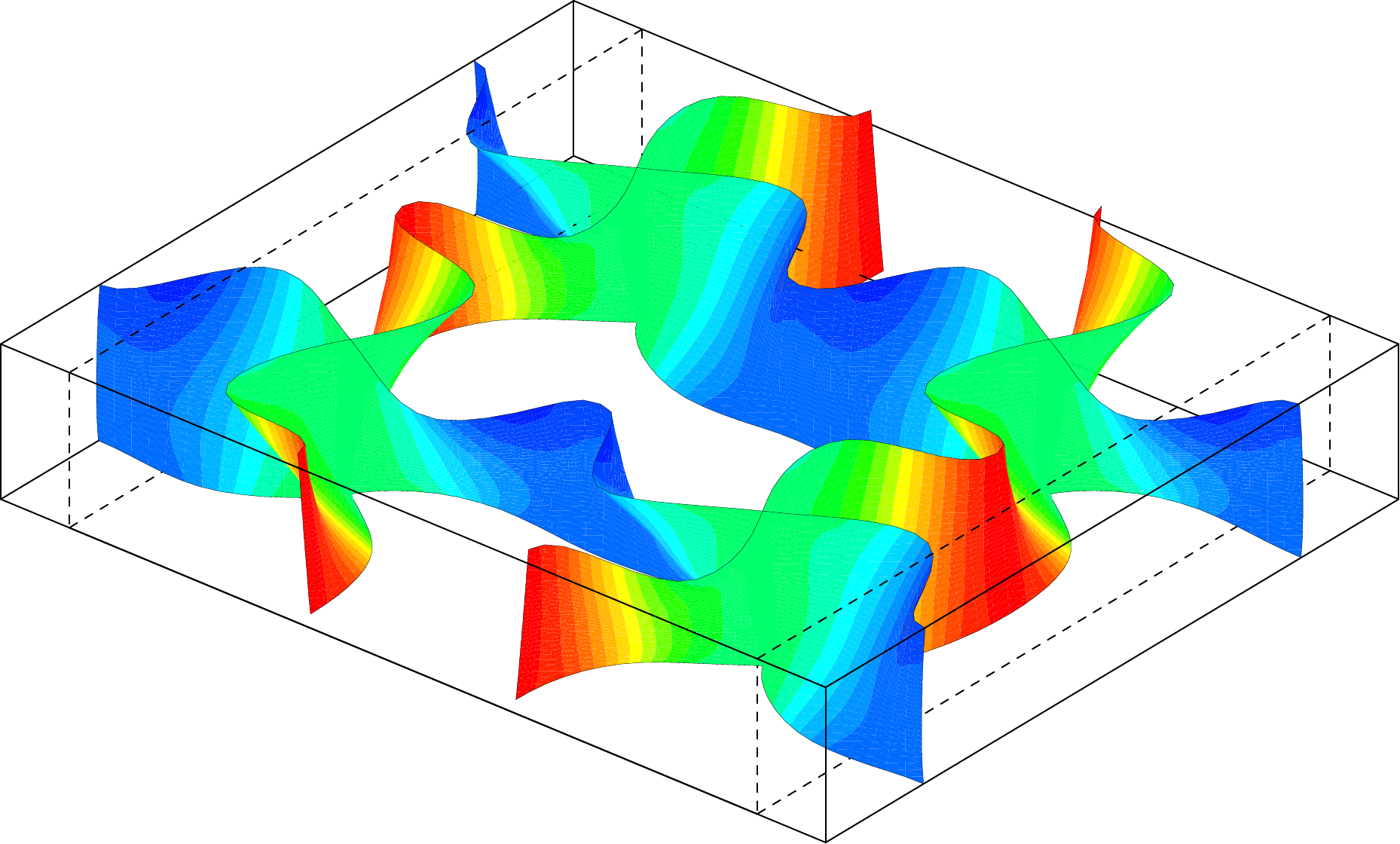}}{(a)}\hfill
\subf{\includegraphics[width=0.25\columnwidth]{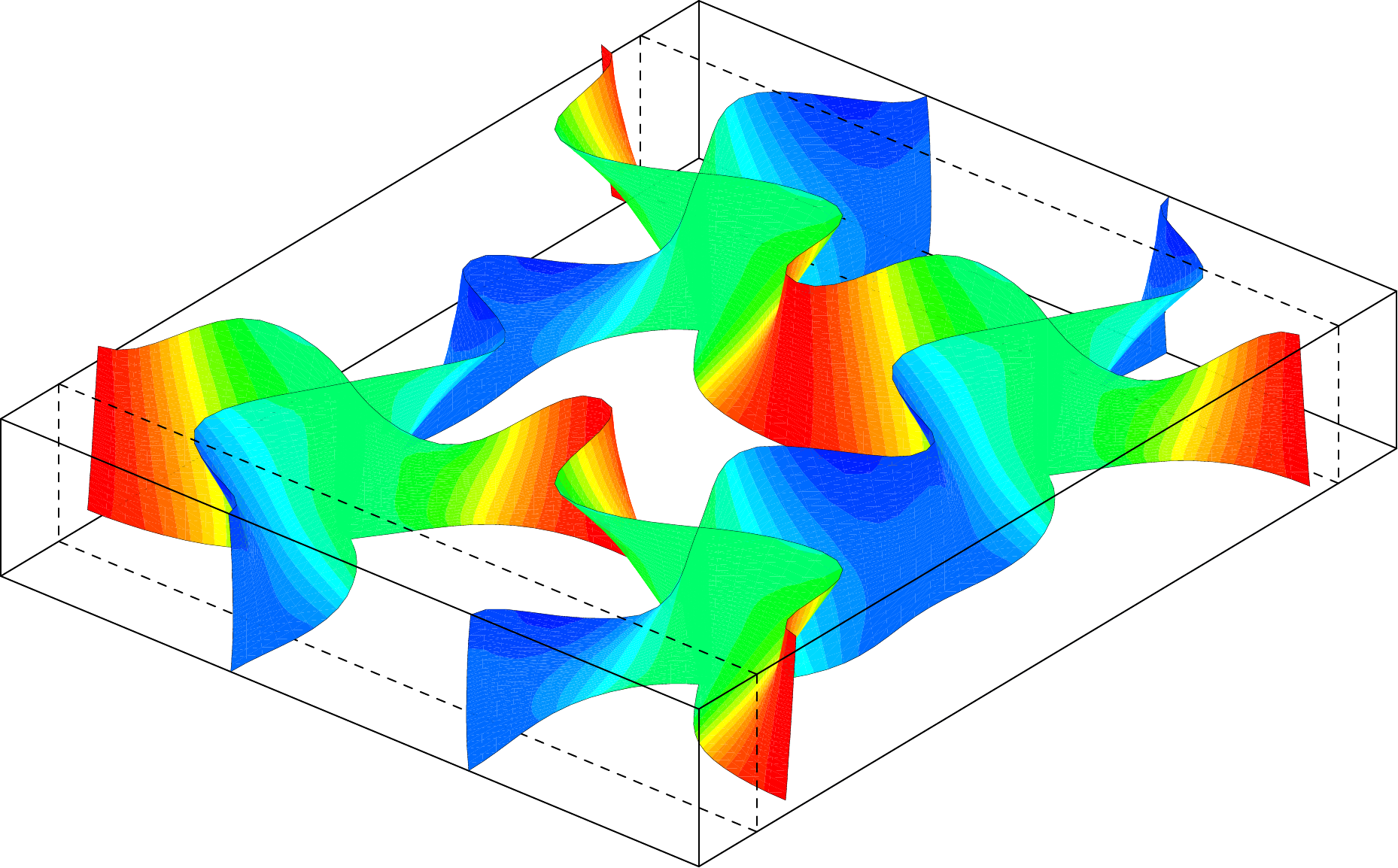}}{(b)}\hfill
\subf{\includegraphics[width=0.25\columnwidth]{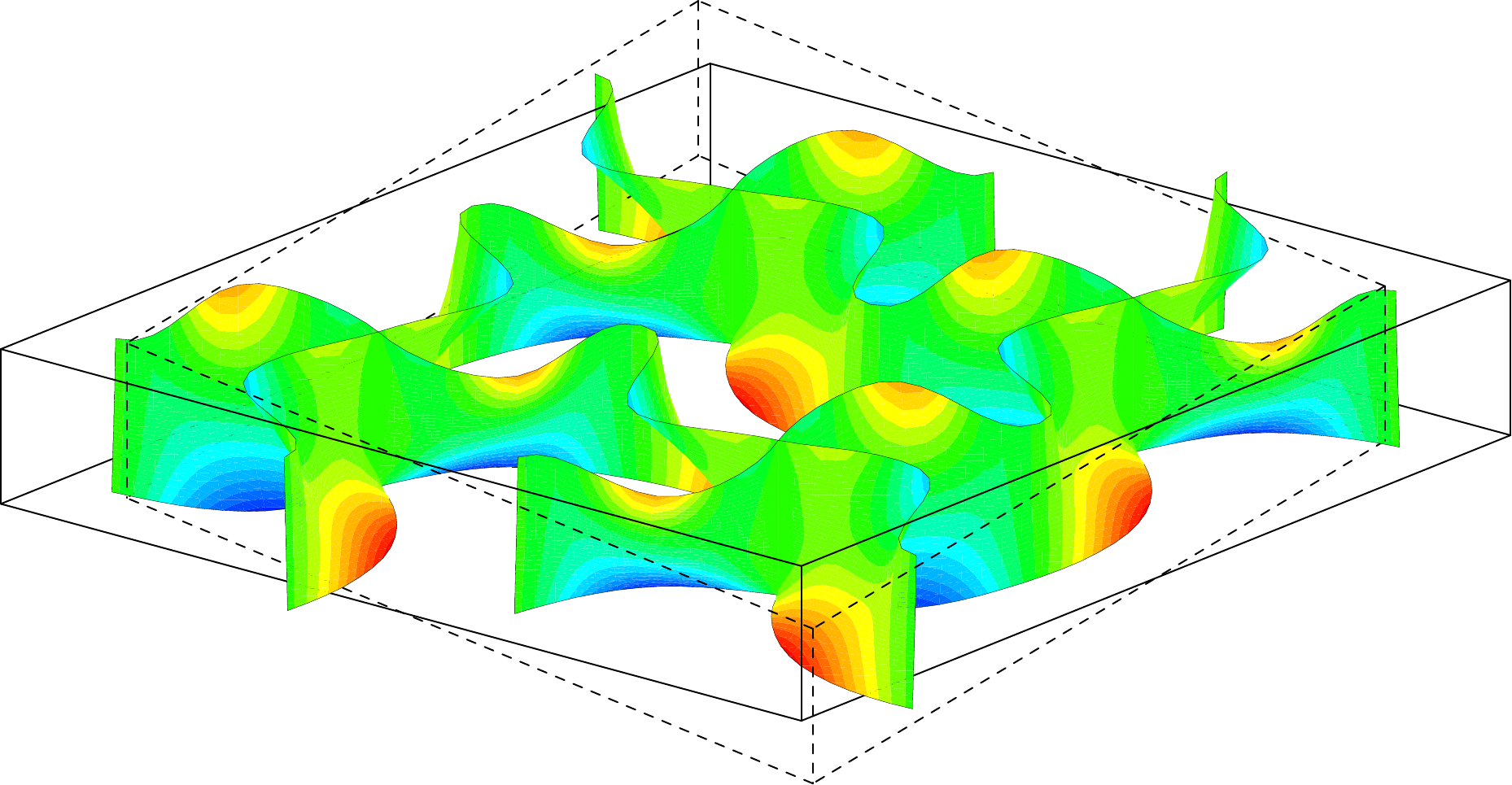}}{(c)}\\[3pt]
\subf{\includegraphics[width=0.25\columnwidth]{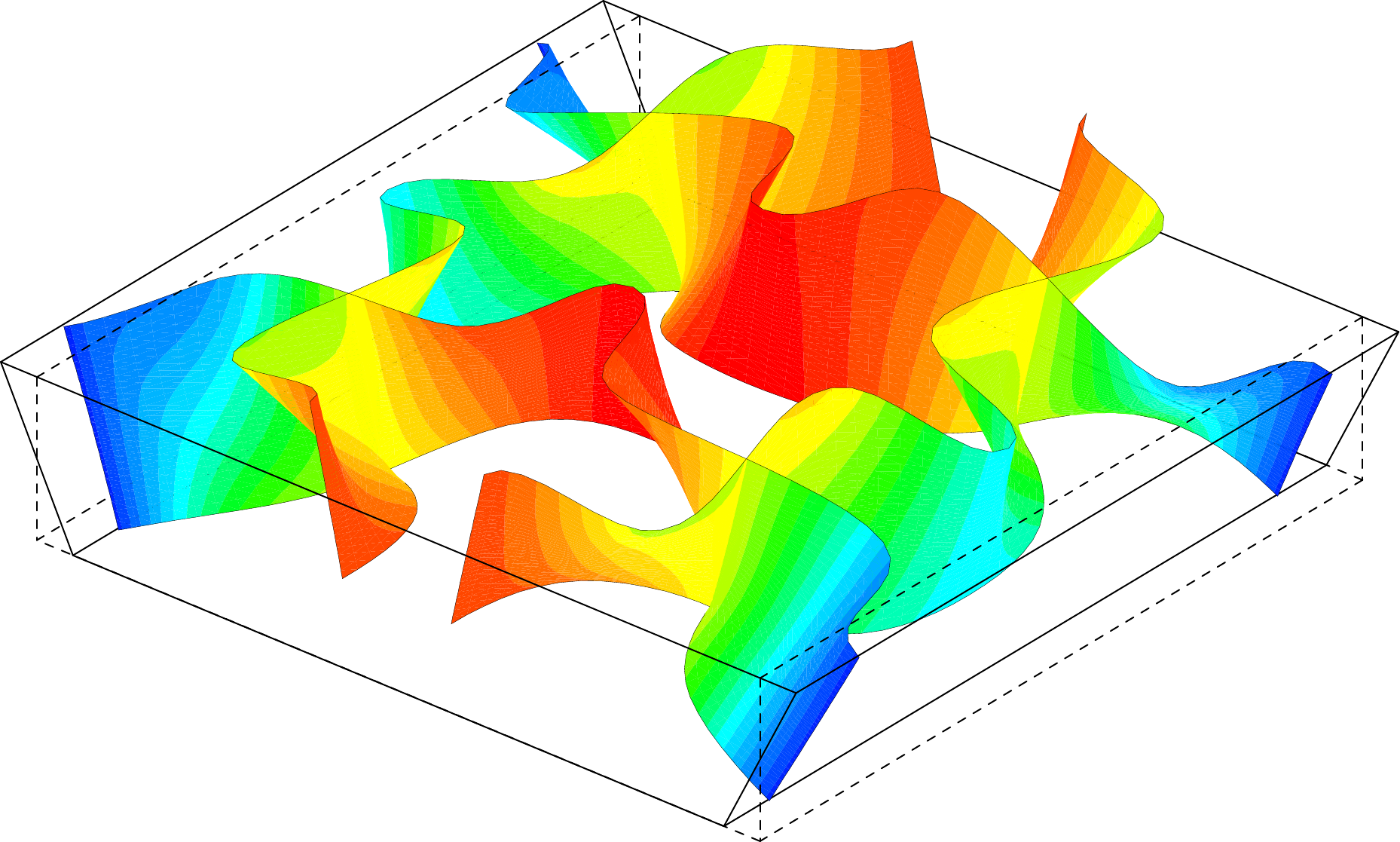}}{(d)}\hfill
\subf{\includegraphics[width=0.25\columnwidth]{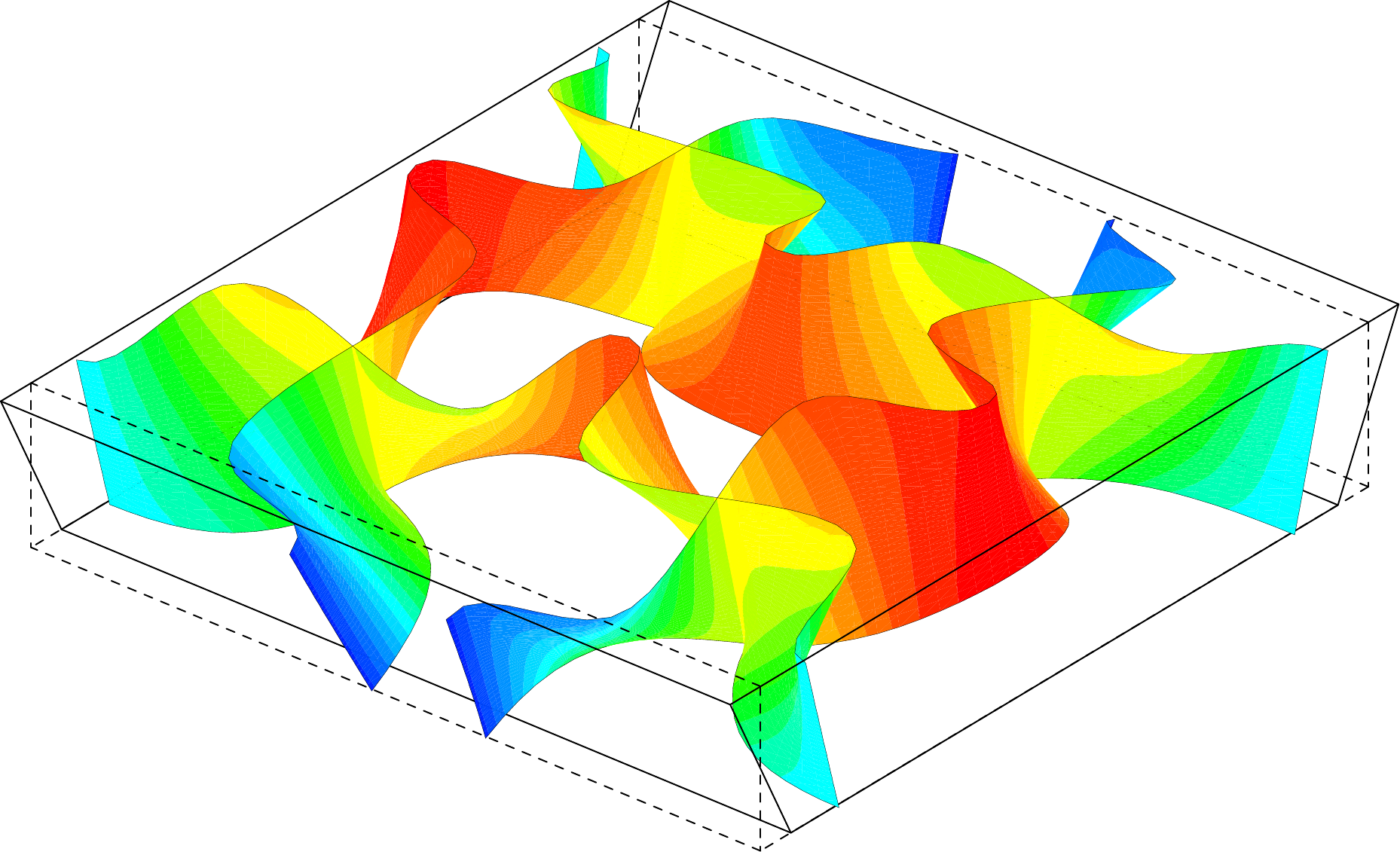}}{(e)}\hfill
\subf{\includegraphics[width=0.25\columnwidth]{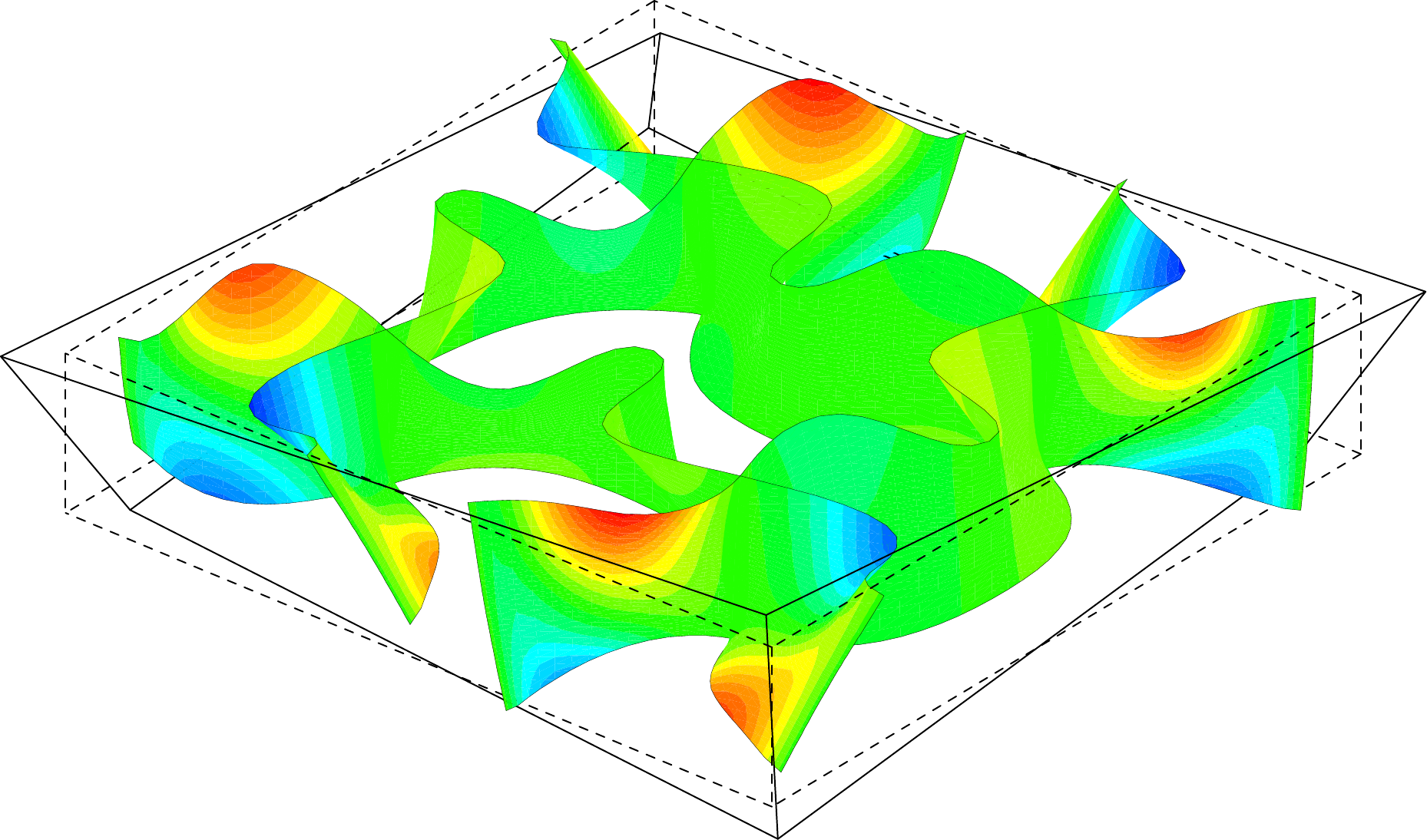}}{(f)}
\caption{Deformation modes of the six solutions of the cells problem \eqref{eq:local-sol-1} and \eqref{eq:local-sol-2}, namely (a-b) two tractions, (c) in-plane shear, (d-e) two  flexures, (f) shear flexure. The colors indicate the ``normalized'' value of the vertical displacement $u_3$ plotted on the deformed mesh. The deformed correspond to a tension up to $20\%$ effective strain}
\label{fig:local-solutions}
\end{figure}

We assume that the material properties can be inhomogeneous but periodic on the microscale. Let the following two-scale asymptotic expansion for the displacement be introduced:
\begin{equation}
\vec{u}^\epsilon (\vec{x}) =
\sum_{\alpha=0}^{+\infty} \epsilon^\alpha \, \vec{u}_\alpha \, (\vec{x},\vec{y}), \quad \vec{y} = \frac{\vec{x}}{\epsilon}.
\label{eq:2scale}
\end{equation}
This leads to a series of problems for different orders of $\epsilon$: at order $\epsilon^{-2}$, we obtain that $u_0(x,y) = u_0(x)$. At order $\epsilon^{-1}$ we obtain the displacement field solutions of the unit cell problems. At order $\epsilon^{0}$ we obtain the linear elastic constitutive equation averaged over the unit cell, yielding the following explicit energy formulation of the homogenised elastic plate tensor $\vec{A}^H$, $\vec{B}^H$ and $\vec{D}^H$ expressed in terms of their cartesian components as:
\begin{equation}
\begin{aligned}
A_{\alpha\beta\gamma\delta}^H =&
\frac{1}{|Y|}\int_{Y} \vec{C}(\vec{y})
\left(\vec{E}^{\alpha\beta}  +\vec{\varepsilon}(\vec{w}^{\alpha\beta}) \right):
\left(\vec{E}^{\gamma\delta} +\vec{\varepsilon}(\vec{w}^{\gamma\delta})\right)
\, d\vec{y},\\
B_{\alpha\beta\gamma\delta}^H =&
\frac{1}{|Y|}\int_{Y} \vec{C}(\vec{y})
\left(\vec{E}^{\alpha\beta}     +\vec{\varepsilon}(\vec{w}^{\alpha\beta})    \right):
\left(y_3\vec{X}^{\gamma\delta} +\vec{\varepsilon}(\vec{\tau}^{\gamma\delta})\right)
\, d\vec{y},\\
D_{\alpha\beta\gamma\delta}^H =&
\frac{1}{|Y|}\int_{Y} \vec{C}(\vec{y})
\left(y_3 \vec{X}^{\alpha\beta}  +\vec{\varepsilon}(\vec{\tau}^{\alpha\beta})\right):
\left(y_3 \vec{X}^{\gamma\delta} +\vec{\varepsilon}(\vec{\tau}^{\gamma\delta})\right)
\, d\vec{y},\\
\end{aligned}
\label{eq:hom_coef}
\end{equation}
where:
\begin{itemize}[leftmargin =*]
\item $\vec{C}$ is the stiffness distribution at the scale of the unit cell.
\item $\vec{E}^{\alpha\beta}$ designates a constant in-plane strain over the unit cell, resulting from the zero order displacement $\vec{u}_0$. There are three independent unit strain fields, namely the horizontal unit strain $\vec{E}^{11} = (1,0,0)^T$, the vertical strain $\vec{E}^{22} = (0,1,0)^T$ and the in-plane shear unit strain $\vec{E}^{12} = (0,0,1)^T$.
\item $\vec{X}^{\alpha\beta}$ designates a constant flexural curvature over the unit cell, resulting from the zero order displacement $\vec{u}_0$. There are three independent unit strain fields, namely the horizontal unit flexure $\vec{X}^{11} = (1,0,0)^T$, the vertical unit flexure $\vec{X}^{22} = (0,1,0)^T$ and the shear unit flexure $\vec{X}^{12} = (0,0,1)^T$.
\item $\vec{w}^{\alpha\beta}$ represents the displacement fields, solution of the following cell problem, expressed in its variational formulation here: 
\begin{equation} 
\begin{cases}
&\text{Find admissible displacement } \vec{w}^{\alpha\beta} \text{ such that} \\[0.1mm]
&\displaystyle \int_Y \vec{C}(\vec{y})(\vec{E}^{\alpha\beta} + \vec{\varepsilon}(\vec{w}^{\alpha\beta})) : \vec{\varepsilon}(\vec{\phi}) \, d\vec{y}= 0 \\[0.1mm]
& \vec{w}^{\alpha\beta} \text{ is } (x_1, x_2)\text{-periodic}.
\end{cases}
\label{eq:local-sol-1}
\end{equation}
where $\vec{\phi}$ are admissible displacement vectors, i.e. with zero mean value and adequate smoothness. The generalized strain components $E^{\alpha\beta}$ are illustrated in \Cref{fig:local-solutions}(a-c).
\item $\vec{\tau}^{\alpha\beta}$ represents the displacement fields, solution of to another cell problem governed by the following equations: 
\begin{equation} 
\begin{cases}
&\text{Find admissible displacement } \vec{\tau}^{\alpha\beta} \text{ such that} \\[0.1mm]
&\displaystyle \int_Y \vec{C}(\vec{y})(y_3 \vec{X}^{\alpha\beta} + \vec{\varepsilon}(\vec{\tau}^{\alpha\beta})) : \vec{\varepsilon}(\vec{\psi}) \, d\vec{y}= 0 \\
& \vec{\tau}^{\alpha\beta} \text{ is } (x_1, x_2)\text{-periodic}.
\end{cases}
\label{eq:local-sol-2}
\end{equation}
where $\vec{\psi}$ are admissible displacement vectors, i.e. with zero mean value and adequate smoothness. The generalized strain components $y_3 X^{\alpha\beta}$ are illustrated in \Cref{fig:local-solutions}(d-f).
\end{itemize} 

In our study, the unit cell is described with shell elements. The numerical computation of the coefficients in  \eqref{eq:hom_coef} is solved numerically as follows:
\begin{equation}
\begin{aligned}
A_{\alpha\beta\gamma\delta}^H =& \frac{1}{|Y|}\int_{\omega} \vec{C}(\vec{y}) 
\left(\vec{E}^{\alpha\beta} + \left(\vec{\mu}(\vec{w}^{\alpha\beta}) + y_3 \vec{\chi}(\vec{w}^{\alpha\beta})\right) \right) : \left(\vec{E}^{\gamma\delta} + \left(\vec{\mu}(\vec{w}^{\gamma\delta})+ y_3 \vec{\chi}(\vec{w}^{\alpha\beta})\right) \right) \, d\vec{y},\\
B_{\alpha\beta\gamma\delta}^H =& \frac{1}{|Y|}\int_{\omega} \vec{C}(\vec{y})
\left(\vec{E}^{\alpha\beta} + \left(\vec{\mu}(\vec{w}^{\alpha\beta}) + y_3 \vec{\chi}(\vec{w}^{\alpha\beta})\right) \right) : \left(y_3\vec{X}^{\gamma\delta} +\left( \vec{\mu}(\vec{\tau}^{\gamma\delta}) + y_3 \vec{\chi}(\vec{\tau}^{\alpha\beta}) \right)\right) \, d\vec{y},\\
D_{\alpha\beta\gamma\delta}^H =& \frac{1}{|Y|}\int_{\omega} \vec{C}(\vec{y})
\left(y_3\vec{X}^{\gamma\delta} +\left( \vec{\mu}(\vec{\tau}^{\gamma\delta}) + y_3 \vec{\chi}(\vec{\tau}^{\alpha\beta}) \right)\right) : \left(y_3\vec{X}^{\gamma\delta} +\left( \vec{\mu}(\vec{\tau}^{\gamma\delta}) + y_3 \vec{\chi}(\vec{\tau}^{\alpha\beta}) \right)\right)
\, d\vec{y},\\
\end{aligned}
\end{equation}
\end{document}